\documentclass[epj,final]{svjour}

\usepackage{graphicx}
\usepackage{dcolumn}
\usepackage{epsfig}
\usepackage{bm}

\begin{document}

\title{Large atom number dual-species magneto-optical trap for fermionic $^6$Li and $^{40}$K atoms}

\author{Armin Ridinger\inst{1}\and Saptarishi Chaudhuri\inst{1}\and Thomas Salez\inst{1}\and Ulrich Eismann\inst{1}\and Diogo Rio Fernandes\inst{1}\and David Wilkowski\inst{2}$^\textrm{\scriptsize,}$\inst{3}\and Frederic Chevy\inst{1}\and Christophe Salomon\inst{1}}

\institute{Laboratoire Kastler Brossel, \'Ecole Normale Sup\'erieure, CNRS, Universit\'e Pierre et Marie-Curie,\\ 24 rue Lhomond, F-75231 Paris Cedex 05, France
\and Centre for Quantum Technologies, National University of Singapore, Singapore 117543, Singapore
\and Institut Non Lin\'eaire de Nice, Universit\'e de Nice Sophia-Antipolis, CNRS, Valbonne 06560, France}
\date{\today}

\abstract{
We present the design, implementation and characterization of a dual-species magneto-optical trap (MOT) for fermionic $^6$Li and $^{40}$K atoms with large atom numbers. The MOT simultaneously contains $5.2\times10^9$ $^6$Li-atoms and $8.0\times10^9$ $^{40}$K-atoms, which are continuously loaded by a Zeeman slower for $^6$Li and a 2D-MOT for $^{40}$K. The atom sources induce capture rates of $1.2\times10^9$ $^6$Li-atoms/s and $1.4\times10^9$ $^{40}$K-atoms/s. Trap losses due to light-induced interspecies collisions of $\sim$65\% were observed and could be minimized to $\sim$10\% by using low magnetic field gradients and low light powers in the repumping light of both atomic species. The described system represents the starting point for the production of a large-atom number quantum degenerate Fermi-Fermi mixture.}
 
\maketitle

\section{Introduction}
The study of ultracold atomic Fermi gases is an emerging research field aiming to understand many-body quantum phenomena occurring in various fields, such as condensed matter systems, disordered systems, quark-gluon plasmas or astrophysics (neutron stars)~\cite{IngKet06,GioPit08}. They provide a unique opportunity to create strongly correlated many-body systems with a high degree of experimental control. One intends to realize analog quantum simulators in Feynman's spirit~\cite{Fey82}, with which many-body Hamiltonians could be solved.

In the field of ultracold Fermi gases the study of mixtures of two different fermionic species with different mass is gaining interest. Both theoretical and experimental aspects motivate this study. Such mixtures are predicted to exhibit a rich phase diagram such as phase separation~\cite{HoCaz09}, crystalline phases~\cite{PetAst07}, exotic pairing mechanisms~\cite{ForGub05} and long-lived trimers~\cite{LevTie09}. They further allow the creation of polar molecules, which have a long-range dipole-dipole interaction~\cite{DeiGro08,NiOsp08}. Two different atomic species yield additional tunable parameters, such as the mass imbalance and species-specific potentials. The mass-imbalance can be varied in an optical lattice, where the effective mass of each species depends on the optical lattice parameters.

The mixture $^6$Li-$^{40}$K is a prime candidate for these studies. $^6$Li and $^{40}$K are the only stable fermionic alkali isotopes and thus belong to the experimentally best-mastered class of atoms. Moreover, both species have \linebreak bosonic isotopes which can also be used to create boson-fermion gases. Furthermore, the mass difference between the two species is large leading to a large electric dipole moment for heteronuclear diatomic molecules (3.6\,D)~\cite{AymDul05}. Finally, many of the above-mentioned predicted quantum phases require strong interspecies interactions and a universal behavior of the gas. It was recently reported~\cite{TieGoo10} that it is possible to reach the universal regime for the $^6$Li-$^{40}$K-mixture due to the existence of a 1.5\,Gauss-wide Feshbach resonance.

The starting point of most mixture experiments is a dual-species magneto-optical trap. It is desirable to capture a large number of atoms at this stage for the following reasons. First, large atom numbers allow to anticipate the losses induced by the subsequent evaporative cooling procedure, which needs to be applied to reach the quantum degenerate regime. Second, a large initial atom number makes the evaporation procedure more efficient. Third, the Fermi temperatures of the gas are larger for larger atom numbers and thus quantum phenomena can be observed at higher temperatures. Finally, a large atom number leads to better signal-to-noise ratios and a greater robustness in day-to-day operation.

A dual-species magneto-optical trap with large atom numbers also allows an efficient creation of ultracold heteronuclear molecules via photoassociation. Using this technique, we have been able to create excited heteronuclear $^6$Li-$^{40}$K* molecules with a formation rate of $\sim5\times10^{7}$s$^{-1}$. The results of this experiment will be the subject of a separate publication~\cite{RidCha11b}.

In this article we describe the design, implementation and characterization of a dual-species magneto-optical \linebreak trap for $^6$Li and $^{40}$K with large atom numbers. In a dual-species MOT, the atom number is in general reduced compared to single-species MOTs due to additional \linebreak interspecies collisions and to experimental constraints, \linebreak such as the imperative to use the same magnetic field for both species or common optics. In other groups working with the $^6$Li-$^{40}$K mixture the following atom numbers have been achieved: in the Munich group~\cite{TagVoi06} the dual-species MOT is loaded from a Zeeman slower for $^6$Li and a vapor for $^{40}$K, resulting in atom numbers of $\sim4\times10^{7}$ ($^6$Li) and $\sim2\times10^{7}$ ($^{40}$K). In the Innsbruck group~\cite{SpiTre10} the dual-species MOT is loaded from a multi-species Zeeman slower and atom numbers of $\sim10^{9}$ ($^6$Li) and $\sim10^{7}$ ($^{40}$K) are achieved. In the group in Amsterdam~\cite{Tie09} two separate 2D-MOTs allow to load $\sim3\times10^{9}$ ($^6$Li) and $\sim2\times10^{9}$ ($^{40}$K). In our setup, the dual-species MOT is loaded from a Zeeman slower for $^6$Li and a 2D-MOT for $^{40}$K. It simultaneously contains $5.2\times10^9$ $^6$Li-atoms and $8.0\times10^9$ $^{40}$K-atoms, which represents a substantial atom number improvement.

For our application in particular a large atom number in the $^{40}$K-MOT is of interest, since we intend to sympathetically cool $^6$Li with $^{40}$K, where $^{40}$K will be prepared and cooled in two different spin states. This approach has been implemented by Tiecke and coworkers~\cite{TieGoo10} and proved to be an efficient cooling method, as it can be realized in a magnetic trap. In this cooling process mostly $^{40}$K-atoms will be lost.	

In future experiments, the atoms stored inside the dual-species MOT will be polarized and magnetically transported to an ultra-high vacuum (UHV) environment with large optical access. There the atom cloud will be evaporatively cooled to quantum degeneracy in an optically plugged magnetic quadrupole trap. Finally it will be transferred into an optical trap to investigate many-body phenomena in lower dimensions. 

This article is organized as follows. In Sec.~\ref{ExperimentalSetup} the experimental setup, including the vacuum assembly and the laser systems, is described. In Sec.~\ref{AtomSources} we present the design and the performance of the atom sources, which are used to load the dual-species MOT, i.e.~a Zeeman slower for $^6$Li and a 2D-MOT for $^{40}$K. In Sec.~\ref{DoubleMOT}, the dual-species MOT is characterized and a study of light-induced interspecies collisions is presented. 

\section{Experimental setup}\label{ExperimentalSetup}

\subsection{Vacuum system}
\begin{figure}[th]
\centering
\includegraphics[width=8cm]{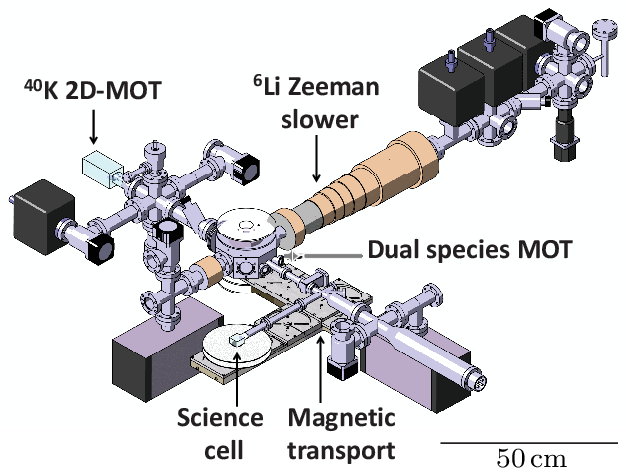}
\caption{(Color online) Schematics of the vacuum assembly. The dual-species MOT is loaded from a 2D-MOT for $^{40}$K and a Zeeman slower for $^6$Li. A magnetic transport allows to transfer the cloud to a UHV science cell with large optical access.}
\label{Vacuum}
\end{figure}

A three-dimensional view of the vacuum system is shown in Fig.~\ref{Vacuum}.
%
%
It consists of two atom trap chambers and three flux regions. The first chamber is a central octagonal chamber where the $^6$Li-$^{40}$K dual-species MOT is prepared. The second chamber is a glass science cell, in which we will evaporatively cool the mixture to quantum degeneracy.

The three flux regions are all connected to the octagonal chamber and are divided in two parts. First, the atom sources, namely a 2D-MOT for $^{40}$K and a Zeeman slower for $^6$Li. Second, a magnetic transport connecting the octagonal chamber to the final science cell. This magnetic transport consists of a spatially fixed assembly of magnetic coils which creates a moving trapping potential of constant shape by applying time-varying currents~\cite{GreBlo01}. It has already been implemented in our system and will be described in a separate publication.

The octagonal chamber can be isolated from the source regions and the science cell by all-metal UHV valves, which allow for separate baking and trouble-shooting. The 2D-MOT and the Zeeman slower region are pumped by one and three 20\,l/s ion pumps, respectively. The octagonal chamber is pumped by a 40\,l/s ion pump and the science chamber by a 40\,l/s ion pump and a titanium sublimation pump. Differential pumping tubes connect the source regions to the octagonal chamber in order to create a high vacuum environment in the octagonal cell. In a similar way, the science chamber is connected to the octagonal chamber via a combination of standard CF16- and homemade vacuum tubes of 1\,cm diameter to further increase the vacuum quality. The glass science cell has a large optical access and permits the installation of an objective for high-resolution imaging.  

\subsection{Laser systems}\label{LaserSystem}
\begin{figure}[th]
\centering
\includegraphics[width=8.75cm]{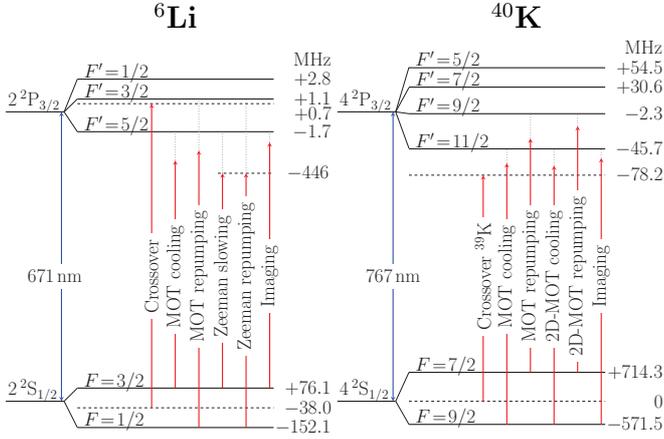}
\caption{(Color online) Level diagrams for the $^6$Li and $^{40}$K D$_2$-lines with their respective hyperfine structures, showing the frequencies required for the dual-species MOT operation. The diode lasers are locked to the indicated saturated absorption crossover signals $2S_{1/2}(F\!=\!1/2,F\!=\!3/2)\rightarrow 2P_{3/2}$ of $^6$Li and $4S_{1/2}(F\!=\!1,F\!=\!2)\rightarrow 4P_{3/2}$ of $^{39}$K.}
\label{Levels}
\end{figure}

\begin{figure}[th]
\centering
\includegraphics[width=8.75cm]{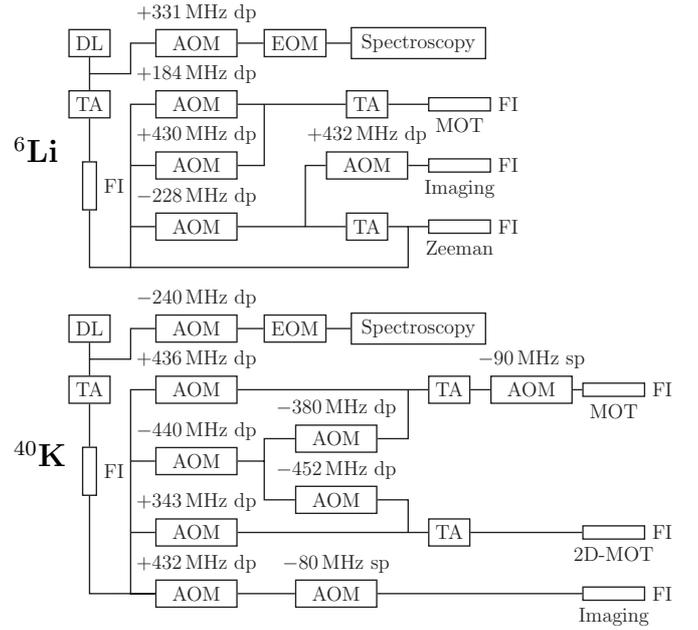}
\caption{Laser systems for $^6$Li and $^{40}$K. The frequencies and amplitudes of the various beams are controlled by AOMs in single pass (sp) or double pass (dp) configuration. The EOMs are used to phase modulate a part of the beam for the diode laser's frequency stabilization. Single mode polarization maintaining fibers (FI) are used for beam shaping and spatial filtering. The indicated AOM frequencies allow to generate the required beam frequencies (see Fig.~\ref{Levels}).}
\label{Optics}
\end{figure}

The dual-species MOT requires separate laser systems and optics for the two different atomic transition wavelengths 671\,nm (Li) and 767\,nm (K). The laser systems provide several beams with different frequencies and intensities for slowing, trapping and probing each atomic species. A sketch of the energy levels of the atomic species and the frequencies of interest are shown in Fig.~\ref{Levels}. The laser systems are set up on separate optical tables and the generated laser beams are transferred to the main experimental table using optical fibers. A simplified scheme of the laser systems is shown in Fig.~\ref{Optics}. Each one consists of a single low output-power frequency-stabilized diode laser (DL) and three tapered amplifiers (TAs) used for light amplification. Due to the small hyperfine splittings of both $^6$Li and $^{40}$K, the required frequencies of the various laser beams are conveniently shifted and independently controlled by acousto-optical modulators (AOMs).
%
%

%
%

The diode lasers are homemade tunable external cavity diode lasers in Littrow configuration. The laser diode for Li (Mitsubishi, ref.~ML101J27) is of low-cost due to its mass production for the DVD industry. Its central free running output wavelength at room temperature is 660\,nm which can be shifted into the range of 671\,nm by heating the diode to 80$^\circ$C. In external cavity configuration its output power is 40\,mW at a driving current of 150\,mA. Under these conditions the laser diode reaches a typical lifetime of 6 months. It can be mode hop-free tuned over a range of 5\,GHz. The laser diode for K is an anti-reflection coated Ridge-Waveguide Laser (Eagleyard, ref.~EYP-RWE-0790-0400-0750-SOT03-0000), whose central free running output wavelength at room temperature corresponds to the desired wavelength. In external cavity configuration its output power is 35\,mW at 90\,mA and it has a typical lifetime of one year. It can be mode hop-free tuned over a range of 10\,GHz.

The tapered amplifiers are commercial semiconductor chips which are mounted on homemade supports. We developed compact support designs with nearly no adjustable parts, which allow for a quick temperature stabilization, do not require running water for heat dissipation and allow for an easy installation process. The support designs are described in detail in the appendix.

We have also developed an all-solid-state laser for \linebreak lithium delivering more than 630\,mW output power, with which we intend to increase further the number of laser-cooled Li atoms. The setup of this light source is described elsewhere~\cite{EisGer11}.

The frequency of each diode laser is stabilized via saturated absorption spectroscopy for which a small part of the DL's output is used (see Fig.~\ref{Optics}). A 20\,MHz electro-optical modulator (EOM) is employed to modulate the phase of the spectroscopy laser beam yielding the derivative of the absorption signal through a lock-in detection. The resulting error signal is transferred to both the diode's current (via a high frequency bias-tee), and, via a PID-controller, to a piezo that adjusts the external cavity's length with a $4$\,kHz bandwidth. An AOM is used to offset the frequency of the diode laser with respect to the absorption line used for locking. It allows for fine adjustments of the frequency while the laser is locked.

The Li diode laser frequency is shifted by $-331$\,MHz from the $^6$Li $2S_{1/2}(F=1/2,F=3/2)\rightarrow 2P_{3/2}$ crossover signal and the K diode laser frequency is shifted by \linebreak $+240$\,MHz from the conveniently located $4S_{1/2}(F=1,F=2)\rightarrow 4P_{3/2}$ crossover signal of $^{39}$K. Note that the small excited state hyperfine structures of both $^6$Li and $^{39}$K are unresolved in the spectroscopy.

The saturated absorption spectroscopy for lithium is realized in a heat pipe of 50\,cm length, in which a natural Li sample (with the isotopic abundances $^7$Li: $92\%$, $^6$Li$: 8\%$) is heated to 350$^\circ$C to create a sufficiently high vapor pressure for absorption. The heat pipe consists of a standard CF40 tube with the Li-sample placed at its center. The tube is heated with a pair of thermocoax cables which are wound around the tube in parallel with opposite current directions in order to prevent magnetic fields to build up. Condensation of lithium atoms on the cell windows needs to be inhibited as Li chemically reacts with glass. This is achieved by adding an argon buffer gas at $\sim0.1$\,mbar pressure, as Ar-Li collisions prevent Li to reach the cell windows in ballistic flight. The optimum argon pressure was chosen such that it provides enough collisions, but does not substantially collision-broaden the absorption spectrum. Water cooling of the metallic parts close to the windows leads to condensation of the diffusing lithium-atoms before those can reach the windows. To avoid that lithium slowly migrates to the colder surfaces, the inside of the tube is covered with a thin stainless steel mesh (Alfa Aesar, ref.~013477), which induces capillary forces acting on the condensed atoms. Since the surface tension of liquid lithium decreases with increasing temperature~\cite{YakMoz00}, the capillary forces cause the atoms to move back to the hotter surfaces.

The saturated absorption spectroscopy for potassium is realized in a cylindrical glass vapor cell of 5\,cm length, in which a natural K-sample (with the isotopic abundances $^{39}$K: $93.36\%$, $^{40}$K: $0.012\%$, $^{41}$K: $6.73\%$) is heated to 40$^\circ$C. Here, a small non-heated appendix of the cell serves as a cold point to prevent condensation of K-atoms on the surfaces crossed by the laser beam.

In both laser systems the frequency stabilized master laser beam is immediately amplified by a first TA and subsequently injected into a single-mode polarization maintaining optical fiber (FI) for beam shaping and spatial filtering (see Fig.~\ref{Optics}). The output beam of the optical fiber is split by a series of polarizing beam splitters into several beams whose frequencies and intensities are independently shifted and controlled with AOMs in single or double pass configuration. The various beams are then recombined with a pair of polarizing beam splitters to linearly polarized bichromatic beams consisting of one cooling and one repumping frequency. Those are then either directly injected into a fiber or into another TA for further amplification. The fibers finally transfer the beams to the main experimental table.

The injection of a bichromatic beam into a TA, whose gain-medium is non-linear, is accompanied with the creation of sidebands~\cite{FerMew99}. The sideband creation is due to parametric amplification of the gain medium by the beating between the two injected frequencies. In general, sidebands represent a loss of the power available in the injected frequencies and can excite unwanted transitions. In our case, where the two injected beam components have significantly different powers and frequencies (differing by $\sim228$\,MHz for $^6$Li and by $\sim1286$\,MHz for $^{40}$K), the power losses are below 10\%. No unwanted transitions are excited by the amplified bichromatic beams, except for the Zeeman slower beam, as that is detuned close to an integer multiple of 228\,MHz and would thus perturb the atoms in the MOT. For this beam the injection of both frequency components into the same TA was thus avoided (see Fig.~\ref{Optics}).

Acoustically isolated homemade mechanical shutters are placed in front of each fiber on the optical tables allowing to switch off the laser beams when required. The shutters consist of a low-cost solenoid-driven mechanical switch (Tyco Electronics, ref.~T90N1D12-12) and a razor blade attached to it via a small rigid lever arm. These shutters typically have a closing time of $\sim100\,\mu$s when placed in the focus of a laser beam and a sufficiently reproducible time delay of the order of 3\,ms.

\section{Atom sources}\label{AtomSources}
Magneto-optical traps can be loaded in different ways. The most efficient is the loading from a beam of slow atoms. This scheme allows isolating the MOT from the atom source region with a differential pumping tube, through which the beam is directed. The MOT thus can be located in a UHV chamber where collisions with the residual gas are minimized. Furthermore, the MOT will be quickly loaded when the atomic beam is cold and has a high flux. The most efficient methods to create such beams are Zeeman slowers and 2D-MOTs. For both atomic species $^6$Li and $^{40}$K, both, Zeeman slowers~\cite{MewFer99,HadGup03,SpiTre10} and 2D-MOTs~\cite{TieGen09}, have been realized in the past. In our setup we chose to implement a Zeeman slower for $^6$Li and a 2D-MOT for $^{40}$K.

\subsection{$^{6}$Li Zeeman slower}
\paragraph{Introduction} Zeeman-tuned slowing represents one of the earliest and most widely used techniques to slow down atoms from an oven~\cite{PhiMet82}. A Zeeman slower longitudinally decelerates an atomic beam using the radiative force of a counter-propagating resonant laser beam. The Doppler effect accumulated during the deceleration is compensated by the Zeeman effect, induced by an inhomogeneous magnetic field, which maintains the atoms on resonance and provides a continuous deceleration.

Two types of Zeeman slowers are commonly used: the positive-field and the sign-changing field (``spin-flip'') Zeeman slower~\cite{MetStr99}. We have implemented a spin-flip Zeeman slower since it brings about several advantages. First, a smaller maximum absolute value of the magnetic field is required. Second, the Zeeman laser beam is non-resonant with the atoms exiting the slower and thus does not push them back into the slower, neither it perturbs the atoms trapped in the $^{6}$Li-MOT. However, the spin-flip Zeeman slower requires repumping light in the region where the magnetic field changes sign and thus makes the optics system slightly more complicated.

\paragraph{Experimental setup} The Zeeman slower consists of two distinct parts: the oven, which creates an atomic beam of thermal atoms, and an assembly of magnetic field coils. In the oven a nearly pure $^{6}$Li sample (5\,g) is heated to 500$\,^\circ$C and an atomic beam is extracted through a collimation tube. The magnetic field coils create an inhomogeneous magnetic field along the flight direction of the atoms.

The oven consists of a vertical reservoir tube (diameter: 16\,mm, length: 180\,mm) and a horizontal collimation tube (diameter: 6\,mm, length: 80\,mm), which is attached to it (see Fig.~\ref{Vacuum}). The upper end of the reservoir tube and the free end of the collimation tube are connected to CF40-flanges. The flange of the reservoir tube is sealed and allows connecting a vacuum pump for baking purposes. The flange of the collimation tube connects the oven to the rest of the vacuum chamber. All parts of the oven are made of stainless steel of type 302L and connected using nickel gaskets instead of copper gaskets as they stand higher temperatures and react less with lithium. The heating of the oven is realized with two high power heating elements (Thermocoax, ref.~SEI 10/50-25/2xCM 10), wound around both, the reservoir and the collimation tube.

The temperature of the oven needs to be stabilized precisely, since the atomic flux critically depends on the temperature. This is accomplished by an active stabilization circuit and an isolation with glass wool and aluminum foil. Along the collimation tube a temperature gradient is maintained in order to recycle lithium atoms sticking to the inner tube walls through capillary action, as explained above. In order to amplify the effect of capillary action, a thin stainless steel mesh with a wire diameter of 0.13\,mm (Alfa Aesar, ref.~013477) is placed inside the tube. This wire decreases the effective diameter of the collimation tube to $\sim5$\,mm. For the operating temperature of 500$\,^\circ$C, the vapor pressure of lithium in the oven amounts to $4\times10^{-3}$\,mbar.

A computer controlled mechanical shutter (Danaher Motion, ref.~BRM-275-03) in front of the oven allows to block the atomic beam during experiments or when the $^{6}$Li-MOT is not in operation.

The oven is pumped through the collimation tube with a 20\,l/s ion pump and isolated from the main chamber via three differential pumping stages and the tube of the Zeeman slower. The pumping efficiency through the collimation tube is $\sim0.19$\,l/s resulting in a pressure drop of a factor $\sim100$. The second and third differential pumping tubes both have a length of 100\,mm and a diameter of 5\,mm and 10\,mm, respectively. A 20\,l/s ion pump is placed after each tube. In total a pressure drop of a factor of $\sim2.5\times10^{6}$ between the oven and the main chamber is obtained.

The assembly of the oven is a three-step procedure. First, the metallic parts of the oven are pre-baked at \linebreak 600$\,^\circ$C during 48\,h. Then, the oven is filled with the lithium sample under air atmosphere and baked again at 600$\,^\circ$C during 12\,h in order to eliminate the impurities in the lithium sample (mostly LiH). Typically 50\% of the sample is lost during this procedure. Then, the oven is connected to the rest of the vacuum chamber under an argon atmosphere, since argon does not react with lithium. Since argon damages ion pumps, the vacuum chamber is first pumped by a turbo molecular pump during 12\,h before the ion pumps are finally launched and the oven is operational.

\begin{figure}[h]
\centering
\includegraphics[width=8.7cm]{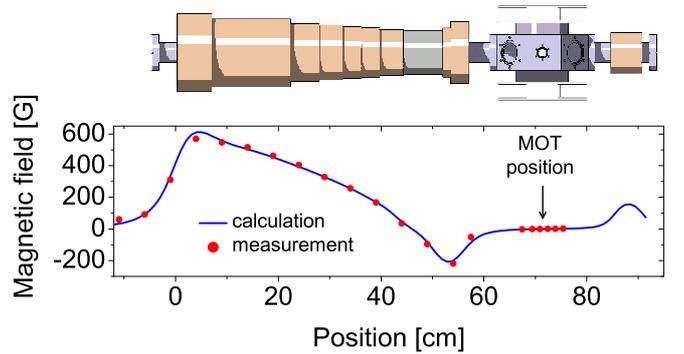}
\caption{(Color online) $^6$Li Zeeman slower coil assembly and generated axial magnetic field profile. The thermal atoms coming from the $^6$Li-oven enter the coil assembly at the position 0, and a fraction of them is slowed down and finally captured in the $^6$Li-MOT, which is located at 71.4\,cm. A compensation coil placed on the opposite side of the MOT (at 84.1\,cm) ensures that the magnetic field is zero at the position of the MOT.}\label{ZEEMANTUBE}
\end{figure}

The Zeeman slower coils are mounted on a 65\,cm long standard CF40 tube placed between the oven and the MOT chamber. A sketch of the coil assembly and the generated axial magnetic field profile are shown in Fig.~\ref{ZEEMANTUBE}. The coil assembly extends over $L=55$\,cm and is separated from the position of the MOT by 16\,cm. The coils are connected in series and were designed such that the desired magnetic field profile is generated for a moderate driving current of 12\,A. The axial magnetic field of the slower along the flight direction of the atoms is measured to be $570$\,G at the entrance and $-220$\,G at the exit. 

%
%
The magnetic field of the Zeeman slower is non-zero at the position of the MOT and hence compensated by a coil placed opposite to the slower coils at a distance of 12.7\,cm from the MOT (see~Fig.~\ref{ZEEMANTUBE}). The compensation coil consists of 4 coil layers wound around a 10\,cm long CF40 standard tube. They are powered by a separate power supply for fine adjustments. When compensated, the magnetic field has an axial gradient of 0.5\,G/cm at the position of the MOT.

The cables of the Zeeman slower coils (APX France, ref.~m\'eplat cuivre \'emaill\'e CL H $1.60\times2.50$) stand bake out procedures up to 200$\,^\circ$C. One layer of a heating cable (Garnisch, ref.~GGCb250-K5-19) is permanently placed underneath the magnetic field coils for these bake out procedures. To avoid heating of the vacuum parts during the Zeeman slower's operation, two layers of water coils were wound underneath the coil layers.

Slowing and repumping light for the Zeeman slower is derived from a bichromatic laser beam which is provided by an optical fiber originating from the laser system. It has a total power of $P_\textrm{\scriptsize fiber}=50$\,mW and its frequencies are both red detuned by $\Delta\omega_\textrm{\scriptsize slow}=\Delta\omega_\textrm{\scriptsize rep}=75\,\Gamma$ from the $2S_{1/2}(F=3/2)\rightarrow2P_{3/2}(F'=5/2)$ slowing and the $2S_{1/2}(F=1/2)\rightarrow2P_{3/2}(F'=3/2)$ repumping transition (see Fig.~\ref{Levels}). The intensity $I_\textrm{\scriptsize slow}$ of the slowing light is 8 times bigger than the intensity $I_\textrm{\scriptsize rep}$ of the repumping light. Both beam components have the same circular polarization ($\sigma^+$ at the position where the atoms enter the slower).

The detuning of the slowing light and the axial magnetic field at the entrance of the coil assembly define the so-called capture velocity $v_\textrm{\scriptsize cap}^\textrm{\scriptsize Zee}$ of the Zeeman slower. All atoms with a velocity smaller than $v_\textrm{\scriptsize cap}^\textrm{\scriptsize Zee}$ are expected to be decelerated to the same final velocity $v_\textrm{\scriptsize fi}^\textrm{\scriptsize Zee}$ at the exit of the slower, provided that they initially populate the correct internal atomic state. The resonance condition for the atoms inside the slower yields $v_\textrm{\scriptsize cap}^\textrm{\scriptsize Zee}\sim830$\,m/s and $v_\textrm{\scriptsize fi}^\textrm{\scriptsize Zee}\sim90$\,m/s. The exit velocity of the slower is thus larger than the capture velocity of the $^{6}$Li-MOT, which is estimated to be $\sim50$\,m/s. However, the atoms are still decelerated significantly in the region between the slower exit and the MOT and are thus expected to be captured by the MOT. The capture velocity of the Zeeman slower is smaller than the most probable thermal speed of the atomic beam, which is given by $v_p=\sqrt{2k_\textrm{\scriptsize B}T/m}=1464$\,m/s at $T=500\,^\circ$C, where $k_\textrm{\scriptsize B}$ denotes the Boltzmann constant and $m$ the mass of the $^{6}$Li-atoms. 

The bichromatic Zeeman slower beam is expanded and focused by a lens pair. The focusing of the beam accounts for the divergence of the atomic beam and the loss of beam power due to absorption and thus yields an efficient utilization of the available laser power. In addition, it induces a small cooling effect along the transverse direction~\cite{MetStr99}. The 1/e$^2$-diameter at the position of the MOT is 31\,mm and the focus is at a distance of 120\,cm from the MOT, 10\,cm behind the oven.  

The divergence of the atomic beam is an important parameter characterizing the Zeeman slower. Three factors contribute to it: first, the geometry of the oven's collimation and the subsequent differential pumping tubes, second the atom's deceleration inside the slower, and third the transverse heating due to the scattered photons during the slowing process. In order to estimate the divergence of the atomic beam, we calculate the maximum possible deflection of an atom which exits the oven with a longitudinal velocity $v_\textrm{\scriptsize cap}^\textrm{\scriptsize Zee}$. An atom with this velocity needs $\sim1.1$\,ms to reach the exit of the Zeeman slower and additional $\sim1.8$\,ms to reach the MOT. Due to the geometry of the collimation and differential pumping tubes it can have a maximum transverse velocity of $\sim$16\,m/s. The change in transverse velocity due to the heating is calculated to be $\sim2.5$\,m/s~\cite{JofKet93} and is thus negligible with respect to the maximum transverse velocity determined by the tube geometry. The final transverse displacement of the atom with respect to the beam axis at the position of the $^{6}$Li-MOT would thus be $\sim5$\,cm, resulting in an effective beam divergence of $\sim$90\,mrad. This divergence requires $^{6}$Li-MOT beams of a large diameter.

\paragraph{Experimental results} For our application the essential parameter which characterizes the performance of the Zeeman slower is the capture rate of the $^{6}$Li-MOT. We studied its dependence as a function of several Zeeman slower parameters, such as: the temperature of the oven, the power of the slowing light, the magnitude of the magnetic field and the intensity ratios between the repumping and slowing light. The optimized values of these parameters are displayed in Tab.~\ref{ZEEMANparameters}, leading to a $^{6}$Li-MOT capture rate of $\sim1.2\times10^{9}$ atoms/s. The capture rate was deduced from a very short loading of the MOT, for which atom losses can still be neglected ($\sim$250\,ms). 
%
\begin{table}[h]
	\centering
		\begin{tabular}{p{5cm} c} 
		  & $^{6}$Li Zeeman slower\\ \hline\hline
			$P_\textrm{\scriptsize fiber}\ [\textrm{mW}]$ & 50\\
			$\Delta\omega_\textrm{\scriptsize slow}\ [\Gamma]$ & -75\\
			$\Delta\omega_\textrm{\scriptsize rep}\ [\Gamma]$ & -75\\
			$I_\textrm{\scriptsize rep}/I_\textrm{\scriptsize slow}$ & 1/8\\
			$B_{\textrm{\scriptsize max}}\ [\textrm{G}]$ & 570\\ \hline\hline
		\end{tabular}
	\caption[3cm]{Optimized values for the parameters of the $^{6}$Li Zeeman slower, yielding a $^{6}$Li-MOT capture rate of $\sim1.2\times10^{9}$ atoms/s at an oven temperature of 500$\,^\circ$C. The definition of the symbols is given in the text. The natural linewidth of $^{6}$Li is $\Gamma/(2\pi)=5.87$\,MHz. The length of the Zeeman slower coil assembly is $55$\,cm.}
	\label{ZEEMANparameters}
\end{table}
%

\begin{figure}[h]
\centering
\includegraphics[width=8.7cm]{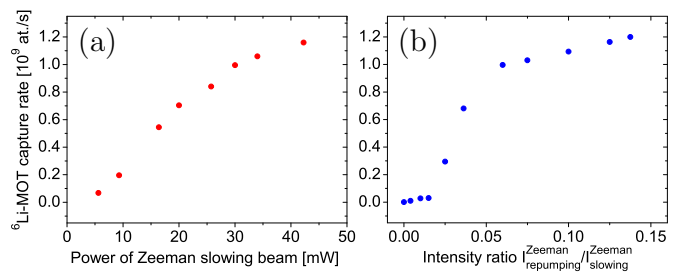}
\caption{(Color online) $^{6}$Li-MOT capture rate as a function of (a) the power of the Zeeman slowing light for a constant repumping light power of 5.6\,mW and (b) the intensity ratio between repumping and slowing light of the Zeeman slower for a constant slowing light power of 45\,mW. The intensities of the superimposed beams depend on the position inside the slower, since the beams are focused toward the oven. At the position where the magnetic field changes sign, a power of 10\,mW corresponds to an intensity of $2.5\,I_\textrm{\scriptsize sat}$, with the saturation intensity $I_\textrm{\scriptsize sat}$ given in Tab.~\ref{MOTparameters}.}\label{ZEEMANFIG1}
\end{figure}

Figure~\ref{ZEEMANFIG1} (a) shows the dependence of the $^{6}$Li-MOT capture rate on the power of the Zeeman slowing light. The curve increases with increasing beam power and indicates saturation for higher powers. In the experiment the slowing light power is 45\,mW, for which the curve in Fig.~\ref{ZEEMANFIG1} (a) starts to saturate, demonstrating that the size of the slowing beam is well chosen. In particular it shows that the beam is not absorbed significantly by the atoms inside the slower.

%
%

The dependence of the $^{6}$Li-MOT capture rate on the intensity ratio between repumping and slowing light of the Zeeman slower is depicted in Fig.~\ref{ZEEMANFIG1} (b). The curve increases with increasing repumping intensity and saturates for higher intensities. For the intensity ratio $I_\textrm{\scriptsize rep}/I_\textrm{\scriptsize slow}\sim0.1$ the repumping intensity in the region where the magnetic field of the Zeeman slower changes sign, is of the order of the saturation intensity. Therefore the transition probability of the repumping transition saturates at $I_\textrm{\scriptsize rep}/I_\textrm{\scriptsize slow}\sim0.1$, explaining the behavior in Fig.~\ref{ZEEMANFIG1} (b). The graph shows that the Zeeman slower only requires a small repumping intensity. It is important that the repumping light has the same circular polarization as the slowing light, since it helps to optically pump the atoms to the cycling transition used for slowing.

\begin{figure}[h]
\centering
\includegraphics[width=8.7cm]{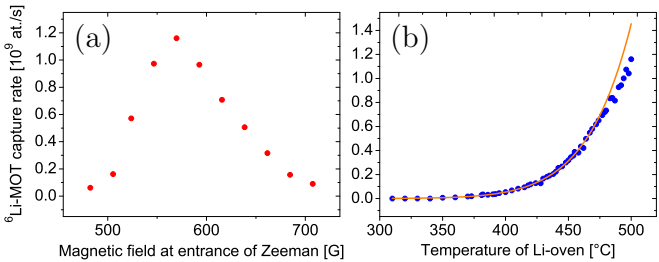}
\caption{(Color online) $^{6}$Li-MOT capture rate as a function of (a) the axial magnetic field of the Zeeman slower and (b) the temperature of the Li-oven. Circles represent the experimental data and the solid curve the theoretical prediction from Eq.~(\ref{MOTflux}).}\label{ZEEMANFIG2}
\end{figure}

Figure~\ref{ZEEMANFIG2} (a) shows the $^{6}$Li-MOT capture rate as a function of the magnitude of the axial magnetic field of the Zeeman slower. The position of the maximum depends on the detuning of the slowing light.

%
%

Figure~\ref{ZEEMANFIG2} (b) shows the dependence of the $^{6}$Li-MOT capture rate on the oven temperature $T$ (circles) as well as a (scaled) theoretical prediction (solid curve) for the experimental data. The curve shows a nearly exponential increase of the capture rate with the temperature. The theoretical prediction is based on a model which assumes no collisions between the atoms (i.e., no intrabeam collisions and no collisions between the beam and the MOT atoms). It is derived as follows.

In the absence of collisions, the normalized velocity distribution of the Zeeman-slowed atoms exiting the slower does not depend on the temperature of the oven. Assuming that the $^{6}$Li-MOT captures mainly atoms which have been slowed by the Zeeman slower, the capture rate $\dot{N}_\textrm{\scriptsize M}$ of the $^{6}$Li-MOT is a temperature-independent fraction of the flux $\dot{N}_\textrm{\scriptsize Z}$ of the Zeeman-slowed atoms: $\dot{N}_\textrm{\scriptsize M}(T)=\kappa_1\dot{N}_\textrm{\scriptsize Z}(T)$. The proportionality constant $\kappa_1$ depends on the divergence of the atomic beam and the capture velocity of the $^{6}$Li-MOT. The flux of the Zeeman-slowed atoms $\dot{N}_\textrm{\scriptsize Z}$ is given by the flux of the oven atoms which have a speed smaller than the Zeeman slower's capture velocity $v_\textrm{\scriptsize cap}^\textrm{\scriptsize Zee}$ and which are in the correct internal atomic state to be decelerated by the Zeeman slower (i.e.~$F=3/2$, $m_F=3/2$). Assuming the oven to be in thermal equilibrium, $\dot{N}_\textrm{\scriptsize Z}$ is given by~\cite{Ram86,TieGen09}
\begin{eqnarray}\label{zeemanflux}
  \dot{N}_\textrm{\scriptsize Z}(T)=\kappa_2n_\textrm{\scriptsize s}(T)A\int_0^{\Omega_\textrm{\tiny Z}}\textrm{d}\Omega\frac{\cos\theta}{4\pi}\int_0^{v_\textrm{\tiny cap}^\textrm{\tiny Zee}}vf(v,T)\textrm{d}v,
\end{eqnarray}
with a temperature-independent constant $\kappa_2$, which equals the fraction of atoms which are in the correct internal atomic state. $n_\textrm{\scriptsize s}(T)$ is the atomic density in the oven, $A=2\times10^{-5}$\,m$^2$ the aperture surface of the oven, $\Omega_\textrm{\scriptsize Z}=A'/l^2=5\times10^{-4}$ the solid angle of the atomic beam (with $A'$ the aperture surface of the last differential pumping tube and $l$ the distance between the two aperture surfaces $A,A'$) and $\textrm{d}\Omega=2\pi\sin\theta \textrm{d}\theta$, with $\theta$ the emission angle with respect to the oven axis. $f(v,T)$ is the normalized speed distribution function given by
\begin{eqnarray}\label{distribution}
  f(v,T)=\sqrt{\frac{2m^3}{\pi k_\textrm{\scriptsize B}^3T^3}}v^2\exp\left(-\frac{mv^2}{2k_\textrm{\scriptsize B}T}\right).
\end{eqnarray}
Since the solid angle of the atomic beam is small, it is $\cos\theta\approx1$ and thus $\int_0^{\Omega_\textrm{\tiny Z}}\textrm{d}\Omega\cos\theta\approx\Omega_\textrm{\scriptsize Z}$.

The explicit temperature dependence of the $^{6}$Li-MOT capture rate is then obtained via $\dot{N}_\textrm{\scriptsize M}(T)=\kappa_1\dot{N}_\textrm{\scriptsize Z}(T)$ by substituting into Eq.~(\ref{zeemanflux}) the ideal gas equation $n_\textrm{\scriptsize s}(T)=p_\textrm{\scriptsize s}/(k_\textrm{\scriptsize B}T)$ and the relation $p_\textrm{\scriptsize s}=p_\textrm{\scriptsize a}\exp[-L_\textrm{\scriptsize 0}/(k_\textrm{\scriptsize B}T)]$ for the saturated vapor pressure $p_\textrm{\scriptsize s}$, with $p_\textrm{\scriptsize a}=1.15\times10^{8}$\,mbar and the latent heat of vaporization $L_\textrm{\scriptsize 0}/k_\textrm{\scriptsize B}=18474$\,K~\cite{AlcItk84}. This relation applies to the temperature range 300-500$\,^\circ$C with an accuracy of 5\%. Thus, we have
\begin{eqnarray}\label{MOTflux}
   \dot{N}_\textrm{\scriptsize M}(T)=\kappa A\Omega_\textrm{\scriptsize Z}p_\textrm{\scriptsize a}\sqrt{\frac{m^3}{8\pi^3k_\textrm{\scriptsize B}^5T^5}}\textrm{e}^{-\frac{L_0}{k_\textrm{\tiny B}T}}\!\!\int_0^{v_\textrm{\tiny cap}^\textrm{\tiny Zee}}\!\!v^3\textrm{e}^{-\frac{mv^2}{2k_\textrm{\tiny B}T}}\textrm{d}v,\quad\!\!
\end{eqnarray}
with $\kappa=\kappa_1\kappa_2$. Scaling Eq.~\ref{MOTflux} to the experimental data for a given (low) temperature ($T\!=\!350^\circ$C) yields the theoretical prediction for the curve shown in Fig.~\ref{ZEEMANFIG2}. The scaling yields $\kappa=10^{-3}$, thus $0.1\%$ of the atoms, which enter the Zeeman slower with a velocity smaller than $v_\textrm{\scriptsize cap}^\textrm{\scriptsize Zee}$, are captured by the $^{6}$Li-MOT.

The main contribution to the small value of $\kappa$ is the large divergence of the slowed atomic beam: $\kappa$ is proportional to the ratio of the atomic beam cross section and the capture surface of the $^{6}$Li-MOT, which is estimated to $\sim10^{-2}$ (assuming the $^{6}$Li-MOT capture surface to be a circle of $1.1\,$cm diameter). Two-dimensional transverse laser cooling of the atomic beam could vastly increase the value of $\kappa$. The remaining $10\%$ are due to an inefficient capture of the $^{6}$Li-MOT and to a significant fraction of oven atoms occupying the incorrect internal atomic states. 

The obtained theoretical prediction agrees well with the experimental data for temperatures below 475$\,^\circ$C (see Fig.~\ref{ZEEMANFIG2} (b)). For temperatures above 475$\,^\circ$C, the experimental data deviate from the prediction indicating that intrabeam collisions or collisions between the atoms in the beam and the MOT become important. We found that for T=500$\,^\circ$C collisions between the thermal $^{6}$Li beam and the trapped $^{6}$Li-MOT atoms indeed take place, which we verified by measuring the lifetime of the $^{6}$Li-MOT in presence and absence of the thermal $^{6}$Li beam, making use of the mechanical block placed at the exit of the oven. The lifetime was found $10\%$ larger for the case where the thermal $^{6}$Li beam was blocked. In a similar way the thermal $^{6}$Li beam also affects the lifetime of the $^{40}$K-MOT. In order to avoid a reduction of the number of trapped $^{40}$K atoms in the dual-species MOT, we therefore limit the $^{6}$Li-oven temperature to 500$\,^\circ$C.

With the help of Eq.~\ref{zeemanflux} the lifetime of the oven can be estimated. Assuming that the collimation tube of the oven recycles all atoms sticking to its wall and the vacuum pumps have no impact on the Li pressure in the oven, the total atomic flux through the collimation tube is obtained by replacing $A'=A$, $v_\textrm{\scriptsize cap}^\textrm{\scriptsize Zee}=\infty$ and $l=8\,$cm (the length of the collimation tube) in Eq.~\ref{zeemanflux}. For the working temperature $T=500\,^\circ$C the lithium vapor pressure is $p_s=4.8\times10^{-3}$\,mbar, corresponding to a density $n_s=4.5\times10^{19}$\,m$^{-3}$. Thus, the atom flux through the collimation tube is $\dot{N}_\textrm{\scriptsize O}=3.5\times10^{14}$\,s$^{-1}\hat{=}3.5\times10^{-12}$\,kg/s. With 3\,g of $^{6}$Li this corresponds to an oven lifetime of $\tau_\textrm{\scriptsize oven}\sim25$\,years. (The importance of the recycling becomes manifest when comparing this value to the hypothetical lifetime of the oven, would the collimation tube be replaced by an aperture of the same surface. In this case the atom flux through this aperture would be $\dot{N}_\textrm{\scriptsize O}^\textrm{\scriptsize hyp}=(\pi l^2/A)\dot{N}_\textrm{\scriptsize O}\sim1000\dot{N}_\textrm{\scriptsize O}$ and thus $\tau_\textrm{\scriptsize oven}^\textrm{\scriptsize hyp}\sim10$\,days.)

\subsection{$^{40}$K 2D-MOT}

\paragraph{Introduction} 2D-MOTs have been widely used over the past years to produce high flux beams of cold atoms~\nocite{DieSpr98,SchBat02,CatMai06,ChaRoy06,TieGen09}\cite{DieSpr98,SchBat02,CatMai06,ChaRoy06,TieGen09,SpiTre10}. In some cases they offer advantages over the more common Zeeman slowers. Even though Zeeman slowers can produce higher fluxes and are more robust, they have the following disadvantages. They produce unwanted magnetic fields close to the MOT which need to be compensated by additional fields, they require a substantial design and construction effort and are space consuming. The atomic beam source of Zeeman slowers needs to be operated at higher temperatures than the vapor cell used as source for 2D-MOTs and the material consumption can be high. In the case of the rare isotope $^{40}$K, this drawback is major: no pure source of $^{40}$K exists and enriched $^{40}$K samples are very expensive (4000\,Euros for 100\,mg of a 4\% enriched sample). Therefore a $^{40}$K Zeeman slower would be very costly. A 2D-MOT can be operated at lower pressures and is thus more economic. In addition it allows separating $^{40}$K from the more abundant $^{39}$K, since it produces an atomic beam which nearly only contains the slowed atoms (i.e.~no thermal background). These considerations motivated us to implement a 2D-MOT for $^{40}$K.

\paragraph{Principle of operation}
In a 2D-MOT, an atomic vapor is cooled and confined transversally and out-coupled longitudinally through an aperture tube. The role of the aperture tube is two-fold. First, it isolates the 2D-MOT from the MOT chamber by differential pumping, and second, it acts as a geometric velocity filter, since only atoms with a small transverse velocity pass through. As the transverse cooling is more efficient for atoms which have a small longitudinal velocity---since those spend more time in the cooling region---most of the transversally cold atoms are also longitudinally cold. Thus, the filter indirectly filters atoms also according to their longitudinal velocity. A 2D-MOT thus produces an atomic beam which is transversally \textit{and} longitudinally cold.

\begin{figure}[h]
\centering
\includegraphics[width=7.25cm]{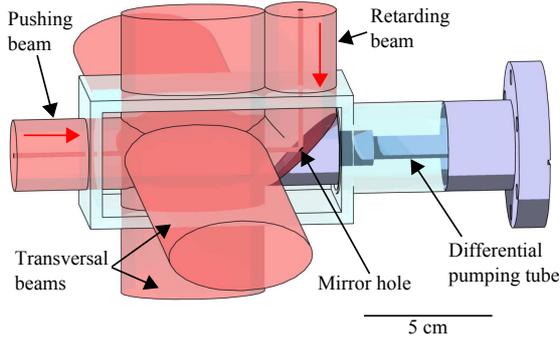}
\caption{(Color online) Sketch of the parallelepipedical glass cell used for the $^{40}$K 2D-MOT. A mirror is placed inside the vacuum chamber to allow an independent control over the longitudinal beam pair. The mirror has a hole in its center and creates a dark cylindrical region in the reflected beams.}\label{2DMOTCELL}
\end{figure}

The flux of a 2D-MOT can be improved by adding a longitudinal molasses cooling to the 2D-MOT configuration~\cite{DieSpr98}. Thus, the atoms spend more time in the transverse cooling region due to the additional longitudinal cooling. The longitudinal beam pair is referred to as the pushing and the retarding beam, where the pushing beam propagates in the direction of the atomic beam (see~Fig.~\ref{2DMOTCELL}). We implemented such a configuration, making use of a $45^{\circ}$-angled mirror inside the vacuum chamber. This mirror has a hole at its center which creates a cylindrical dark region in the reflected retarding beam. In this region, the atoms are accelerated along the longitudinal direction by the pushing beam only, which allows an efficient out-coupling of the atomic beam.

\paragraph{Experimental setup}
The vacuum chamber of the \linebreak 2D-MOT consists of standard CF40 components and a parallelepipedical glass cell (dimensions 110\,mm$\times$55\,mm$\times$\linebreak55\,mm), which is depicted in Fig.~\ref{2DMOTCELL}. Its long axis is aligned horizontally, parallel to the differential pumping tube and the direction of the produced atomic beam. The mirror inside the vacuum chamber is a polished stainless steel mirror with an elliptical surface (diameters 3.0\,cm and 4.2\,cm). It is attached to the differential pumping tube inside the vacuum. It allows to overlap the two longitudinal laser beams whose powers and orientations can thus be independently controlled externally. The mirror's material has a reflectivity of only $50\%$, but inhibits chemical reaction of potassium with its surface. The differential pumping tube intercepts the mirror at its center. The tube has a diameter of 2\,mm over a distance of 1.5\,cm and then stepwise widens up to 10\,mm over a total distance of 22\,cm. The $^{40}$K-MOT is located 55\,cm away from the 2D-MOT center. Assuming a ballistic flight of the atoms, the geometry of the differential pumping tube defines an upper limit of the divergence of the atomic beam, which is calculated to be $\sim$35\,mrad. The atomic beam thus is expected to have a diameter of $\sim2$\,cm when it reaches the $^{40}$K-MOT. The differential pumping tube has a conductance of 0.04\,l/s. The generated pressure ratio between the 2D-MOT and the 3D-MOT chambers is $\sim10^3$. 

%
%
The potassium source is an isotopically enriched $^{40}$K sample (containing 4\,mg of $^{40}$K, 89.5\,mg of $^{39}$K and 6.5\,mg of $^{41}$K, from Technical Glass Inc., Aurora, USA), placed at a distance of 20\,cm from the glass cell. It was purchased in a small ampule which was broken under vacuum inside a modified stainless steel CF16 bellow. The small vapor pressure of potassium at room temperature ($10^{-8}$\,mbar) requires heating of the entire 2D-MOT chamber. We heat the source region to 100$^\circ$C, all intermediate parts to 80$^\circ$C and the glass cell to 45$^\circ$C. The gradient in temperature ensures that the potassium migrates into the cell and remains there. The resulting K-pressure in the glass cell was measured by absorption of a low intensity probe. We found $2.3\times10^{-7}$\,mbar, which implies a partial pressure of the $^{40}$K-isotope of $1\times10^{-8}$\,mbar. In contrast to lithium, the source lifetime is mainly determined by the pumping speed of the ion pump. At the measured pressure the lifetime of the source is estimated to $\sim2$ years.

Four air-cooled rectangular shaped elongated racetrack coils (dimensions 160\,mm$\times$60\,mm) are placed around the glass cell to produce a 2D quadrupole field with cylindrical symmetry and a horizontal line of zero magnetic field. This racetrack coil geometry allows an independent control of the transverse position of the magnetic field zero, and minimizes finite coil fringe effects at the coil ends. The coils are controlled by four separate power supplies. For optimized operation, the transverse magnetic field gradients are $\partial_x B=\partial_y B=11\,$G/cm.

Cooling and repumping light for the 2D-MOT is derived from a bichromatic laser beam which is provided by an optical fiber originating from the laser system. It has a total power of $P_\textrm{\scriptsize fiber}=450$\,mW and its frequencies are red detuned by $\sim3.5\,\Gamma$ from the $4S_{1/2}(F=9/2)\rightarrow4P_{3/2}(F'=11/2)$ cooling and by $\sim2.5\,\Gamma$ from the $4S_{1/2}(F=7/2)\rightarrow4P_{3/2}(F'=9/2)$ repumping transition (see Fig.~\ref{Levels}). The beam is separated into four beams and expanded by spherical and cylindrical telescopes to create the transverse and longitudinal 2D-MOT beams. The transverse beams have an elliptical cross section (1/e$^2$-diameters: 27.5\,mm and 55\,mm), are circularly polarized and retro-reflected by right-angled prisms, which preserve the helicity of the beams. The power losses in the surface of the glass cell and the prisms weaken the power of the retro-reflected beams by $\sim17$\% (the loss contribution of the absorption by the vapor is negligible due to the high laser power). This power imbalance is compensated by shifting the position of the magnetic field zero. The longitudinal beams are linearly polarized and have a circular cross section (1/e$^2$-diameter: 27.5\,mm). $75\%$ of the fiber output power is used for the transverse beams, $25\%$ for the longitudinal beams. The intensity ratio between pushing and retarding beam along the atomic beam axis is $\sim6$ (for reasons explained below). 

\paragraph{Experimental results}
For our purpose the essential parameter which characterizes the performance of the 2D-MOT is the capture rate of the $^{40}$K-MOT. We studied its dependence as a function of several 2D-MOT parameters, such as: the vapor pressure in the 2D-MOT cell, the total cooling light power, the detuning of the cooling frequency and the intensity ratios between the repumping and cooling light and between the pushing and retarding beam. The optimized values of these parameters are displayed in Tab.~\ref{2DMOTparameters}, leading to a $^{40}$K-MOT capture rate of $\sim1.4\times10^{9}$ atoms/s.
%
\begin{table}[h]
	\centering
		\begin{tabular}{p{5cm} c} 
		  & $^{40}$K 2D-MOT\\ \hline\hline
			$P_\textrm{\scriptsize fiber}\ [\textrm{mW}]$ & 450\\
			$\Delta\omega_\textrm{\scriptsize cool}\ [\Gamma]$ & -3.5\\
			$\Delta\omega_\textrm{\scriptsize rep}\ [\Gamma]$ & -2.5\\
			$I_\textrm{\scriptsize rep}/I_\textrm{\scriptsize cool}$ & 1/2\\
			$I_\textrm{\scriptsize push}/I_\textrm{\scriptsize ret}$ & 6\\
			$\partial_x B, \partial_y B\ [\textrm{G/cm}]$ & 11\\
			K vapor pressure\ [mbar] & $2.3\times10^{-7}$\\\hline\hline
		\end{tabular}
	\caption[3cm]{Optimized values for the parameters of the $^{40}$K 2D-MOT, yielding a $^{40}$K-MOT capture rate of $\sim1.4\times10^{9}$ atoms/s. The definition of the symbols is given in the text. The natural linewidth of $^{40}$K is $\Gamma/(2\pi)=6.04$\,MHz.}
	\label{2DMOTparameters}
\end{table}
%

The mean velocity of the atoms in the atomic beam can be estimated as follows. It is approximately given by the average time required for the atoms of the 2D-MOT region to reach the 3D-MOT. This time was measured by recording the time delay of the onset of the $^{40}$K-MOT loading after switching on the 2D-MOT beams. We measured a time delay of $\sim$23\,ms and deduce a mean longitudinal velocity of the captured atoms of $\sim$24\,m/s. At this velocity, the displacement due to gravity of the beam of atoms from the $^{40}$K-MOT center is $\sim$2.6\,mm, which is negligible compared to the size of the $^{40}$K-MOT beams and the divergence of the atomic beam.

\begin{figure}[h]
\centering
\includegraphics[width=8.7cm]{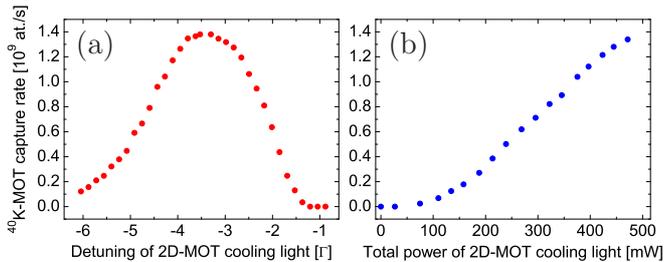}
\caption{(Color online) $^{40}$K-MOT capture rate as a function of (a) the detuning and (b) the total power of the cooling light used for the 2D-MOT (for a constant intensity ratio between the cooling and repumping light). The total power refers to the sum of the powers in the six 2D-MOT beams, where a power of 470\,mW corresponds to a total intensity of $\sim47$\,$I_\textrm{\scriptsize sat}$ at the center of the 2D-MOT, with the saturation intensity $I_\textrm{\scriptsize sat}$ given in Tab.~\ref{MOTparameters}.}\label{2DMOTFIG1}
\end{figure}

Figure~\ref{2DMOTFIG1} (a) shows the dependence of the $^{40}$K-MOT capture rate on the detuning $\Delta\omega_\textrm{\scriptsize cool}$ of the 2D-MOT cooling light. The curve has a maximum at $\Delta\omega_\textrm{\scriptsize cool}=-3.5\,\Gamma$ and a full width at half maximum (FWHM) of $2.7\,\Gamma$. The maximum is the result of two opposing effects: the scattering force of the 2D-MOT beams decreases with increasing detuning whereas the capture velocity increases~\cite{MetStr99}. The first effect implies a less efficient transverse cooling whereas the second leads to a more efficient capture of atoms. An additional effect might influence the shape of the curve: since the scattering force of the pushing beam depends on the detuning, also the mean-velocity of the atomic beam depends on it~\nocite{DieSpr98,SchBat02}\cite{DieSpr98,SchBat02,ChaRoy06}. 
Since we measure the $^{40}$K-MOT capture rate rather than the flux of the 2D-MOT, the mean-velocity might exceed the capture velocity of the $^{40}$K-MOT. However, as shown in refs.~\nocite{DieSpr98,SchBat02}\cite{DieSpr98,SchBat02,ChaRoy06}, the mean-velocity of the beam only slightly changes with the detuning, such that we expect this effect to only weakly influence the curve. From the shape of the curve we conclude that the $^{40}$K-MOT capture rate is not very sensitive to changes of $\Delta\omega_\textrm{\scriptsize cool}$. 

%
%

The dependence of the $^{40}$K-MOT capture rate on the total power of the 2D-MOT cooling light is depicted in Fig.~\ref{2DMOTFIG1} (b). The total power refers to the sum of the powers in the six 2D-MOT beams. According to the chosen beam sizes, the maximum power of 470\,mW corresponds to a total intensity of $\sim47$\,$I_\textrm{\scriptsize sat}$ (for zero detuning) at the center of the 2D-MOT, with the saturation intensity $I_\textrm{\scriptsize sat}$ given in Tab.~\ref{MOTparameters}. The curve almost linearly increases with light power without a clear indication of saturation. The increase is due to two effects. First, the 2D-MOT capture velocity increases with laser power due to the power broadening of the atomic spectral lines. Second, the scattering force increases, resulting in a steeper transverse confinement, which facilitates the injection of the atoms into the differential pumping tube. At some point, the curve is expected to saturate, since the temperature of the cooled atoms and light-induced collisions between them increase with light power. These effects, however, are less limiting in a 2D-MOT as compared to a 3D-MOT, since the atomic density in a 2D-MOT is typically three orders of magnitude smaller due to the absence of a three-dimensional confinement. Thus, in a 2D-MOT a high light power would be required to reach the regime of saturation.

\begin{figure}[h]
\centering
\includegraphics[width=8.7cm]{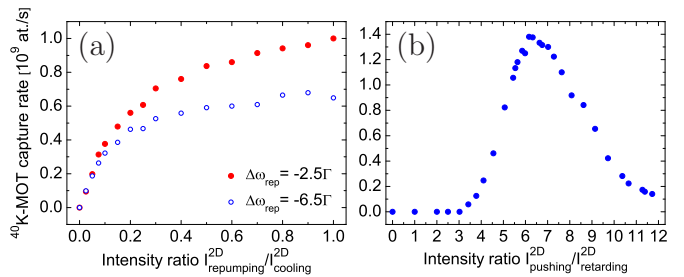}
\caption{(Color online) $^{40}$K-MOT capture rate as a function of the intensity ratio between (a) repumping and cooling light of the 2D-MOT for two different repumping detunings $\Delta\omega_\textrm{\scriptsize rep}$ and a constant total cooling light power of 300\,mW (which corresponds to a total intensity of $\sim30$\,$I_\textrm{\scriptsize sat}$) and (b) the pushing and the retarding beams of the 2D-MOT. The intensities of the pushing and retarding beams refer to the intensities along the atomic beam axis.}\label{2DMOTFIG2}
\end{figure}

Figure~\ref{2DMOTFIG2} (a) shows the dependence of the $^{40}$K-MOT capture rate on the intensity ratio between the cooling and repumping light of the 2D-MOT for the two different repumping detunings $\Delta\omega_\textrm{\scriptsize rep}^{(1)}=-2.5\,\Gamma$ and $\Delta\omega_\textrm{\scriptsize rep}^{(2)}=-6.5\,\Gamma$ and for a constant total cooling light power of 300\,mW. The graph shows that for both frequencies the $^{40}$K-MOT capture rate increases with increasing repumping intensity and that it saturates at high intensities. It also shows that the maximum capture rate is bigger for the smaller detuning. The intensity dependence of the curves results from the likewise intensity dependence of the transition probability for an atomic transition. The maximum capture rate is bigger for the smaller detuning, since this detuning contributes more efficiently to the cooling process. In our experiment, a fixed total laser power is available for both repumping and cooling light. It is distributed such that the resulting capture rate is maximized. It was found to be maximum for an intensity ratio of $I_\textrm{\scriptsize rep}/I_\textrm{\scriptsize cool}\sim1/2$. For that ratio the detuning $\Delta\omega_\textrm{\scriptsize rep}^{(2)}=-2.5\,\Gamma$ also yields the maximum capture rate.

%
%

The dependence of the $^{40}$K-MOT capture rate on the intensity ratio between pushing and retarding beam is depicted in Fig.~\ref{2DMOTFIG2} (b). The curve has a maximum at \linebreak $I_\textrm{\scriptsize push}/I_\textrm{\scriptsize retard}\sim6$. It is zero for values of $I_\textrm{\scriptsize push}/I_\textrm{\scriptsize retard}$ between 0 and 3, then increases until the maximum and falls off again with a smaller slope. From the curve we can extract information about the importance of the reflectivity of the mirror inside the vacuum and of the size of its hole. For a given intensity ratio $I_\textrm{\scriptsize push}/I_\textrm{\scriptsize retard}$ along the (horizontal) direction of the atomic beam, the mirror's reflectivity determines the intensity ratio $I_\textrm{\scriptsize push}^*/I_\textrm{\scriptsize retard}^*$ along the vertical direction above the reflecting surface of the mirror (see Fig.~\ref{2DMOTCELL}). If $I_\textrm{\scriptsize push}^*/I_\textrm{\scriptsize retard}^*$ differs from 1, the atomic beam can experience a vertical deflection in this region. The hole inside the mirror creates a dark cylinder in the pushing beam after its reflection, so that in the region above the hole only light from the retarding beam has a vertical direction, which can also give rise to a vertical deflection of the atomic beam.

In the following we estimate the deflection of the atomic beam, which is induced by the unbalanced retarding beam in the small region above the hole. Assuming the atomic beam to have reached its final longitudinal velocity of 24\,m/s when entering into the hole, the atoms spend 85\,$\mu$s in the region above the hole. Neglecting Doppler shifts and the presence of the pushing beam along the horizontal direction (no transverse beams are present in the region above the mirror), the atoms will scatter $N_\textrm{\scriptsize ph}=R_\textrm{\scriptsize sc}\times(85\,\mu s)\sim75$ photons, with $R_\textrm{\scriptsize sc}$ being the scattering rate~\cite{MetStr99} for the given detuning $\Delta\omega_\textrm{\scriptsize cool}=-3.5\,\Gamma$ and peak intensity $I_\textrm{\scriptsize retard}^*=2.5I_\textrm{\scriptsize sat}$. The recoil velocity of $^{40}$K being given by $v_\textrm{\scriptsize rec}=0.013$\,m/s, each atom will accumulate a transverse velocity of $v_\textrm{\scriptsize dev}\sim1$\,m/s. This leads to a downwards deflection of the atomic beam by an angle of $\sim40\,$mrad, which is more than a factor two bigger than the maximum deflection angle allowed by the differential pumping tubes. The atoms will thus not reach the $^{40}$K-MOT.

This deflection needs to be anticipated by an intensity imbalance $I_\textrm{\scriptsize push}^*>I_\textrm{\scriptsize retard}^*$ in the region above the reflecting surface of the mirror, as that results in an upwards deflection of the atomic beam. For the given mirror reflectivity of $50\%$, $I_\textrm{\scriptsize push}^*>I_\textrm{\scriptsize retard}^*$ is equivalent to $I_\textrm{\scriptsize push}/I_\textrm{\scriptsize retard}>4$, which corresponds to the experimental observation depicted in Fig.~\ref{2DMOTFIG2} (b). The deflection of the atomic beam in the region above the hole could be avoided using a beam block which creates a dark cylinder in the region above the mirror which overlaps with the one in the pushing beam. In this configuration the position of the curve optimum in Fig.~\ref{2DMOTFIG2} (b) would change from $I_\textrm{\scriptsize push}/I_\textrm{\scriptsize retard}=6$ to $I_\textrm{\scriptsize push}/I_\textrm{\scriptsize retard}=4$. For mirrors with a reflectivity close to 100\% the position of the curve optimum could thus even be changed to $I_\textrm{\scriptsize push}/I_\textrm{\scriptsize retard}=1$, for which the longitudinal optical molasses cooling would be most efficient leading to a maximum 2D-MOT flux. Due to the polarization gradients generated by the transverse 2D-MOT beams the longitudinal optical molasses cooling is, however, still very efficient even in case of an intensity imbalance of 6 along the atomic beam axis. 

\begin{figure}[h]
\centering
\includegraphics[width=8.7cm]{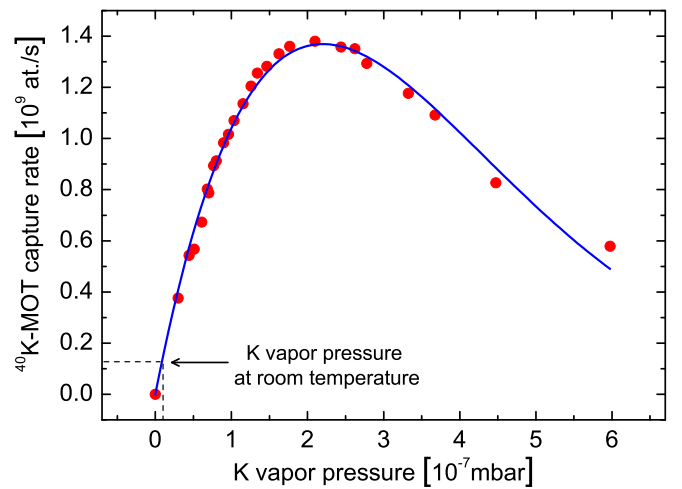}
\caption{(Color online) $^{40}$K-MOT capture rate as a function of the potassium vapor pressure (all isotopes). Circles: experimental data, solid curve: fit of the experimental data by Eq.~(\ref{pressurefit2}). Due to the low abundance of the $^{40}$K-isotope in our potassium sample (4\%), the $^{40}$K-MOT capture rate is limited by collisions between the $^{40}$K-atoms and the other K-isotopes in the 2D-MOT cell. At room temperature the potassium vapor pressure is $1\times10^{-8}$\,mbar.}\label{2DMOTFIG3}
\end{figure}

We now study the dependence of the $^{40}$K-MOT capture rate on the vapor pressure of potassium (all isotopes) in the 2D-MOT cell, which is shown in Fig.~\ref{2DMOTFIG3} (circles) together with a fit to a theoretical model (solid curve). The vapor pressure was measured by recording the absorption profile of a low intensity probe. The curve in Fig.~\ref{2DMOTFIG3} has a maximum at a vapor pressure of $2.3\times10^{-7}$\,mbar. In the absence of collisions, the curve should increase linearly with pressure, which is indeed observed for low pressures. For high pressures, collisions become important and limit the $^{40}$K-MOT capture rate. The dependence of the $^{40}$K-MOT capture rate $L$ on the pressure $p$ can be described by the function~\cite{ChaRoy06}
\begin{eqnarray}\label{pressurefit}
L=L_0\exp\left[-\left(\Gamma_\textrm{\scriptsize coll}+\beta\int n^2(\textbf{r})d^3r\right)\left\langle t_\textrm{\scriptsize cool}\right\rangle\right],
\end{eqnarray}
where $L_0$ denotes the hypothetical capture rate of the $^{40}$K-MOT in the absence of collisions in the 2D-MOT chamber, $\Gamma_\textrm{\scriptsize coll}$ denotes the collisional loss rate due to collisions in the 2D-MOT chamber between the cooled atoms and the background atoms, $\left\langle t_\textrm{\scriptsize cool}\right\rangle$ is the average time which the atoms spend inside the 2D-MOT cooling region, $n(\textbf{r})$ is the position-dependent atomic density in the atomic beam, and $\beta$ is the two-body loss rate coefficient which describes the cold collisions between the $^{40}$K atoms in the atomic beam. $L_0$ is proportional to the atomic density $n_\textrm{\scriptsize K}$ in the vapor cell, and $\Gamma_\textrm{\scriptsize coll}=n_\textrm{\scriptsize K}\sigma_\textrm{\scriptsize eff}\left\langle v\right\rangle$, where $\sigma_\textrm{\scriptsize eff}$ is the effective collision cross section, and $\left\langle v\right\rangle\sim400$\,m/s the mean velocity of the thermal potassium atoms. The term describing the cold collisions is approximately proportional to $n_\textrm{\scriptsize K}^2$ due to the small density obtained in the 2D-MOT. For the investigated pressure range, the ratio $p/n_\textrm{\scriptsize K}$ only changes slightly with temperature and can thus be considered constant. Therefore Eq.(\ref{pressurefit}) can be written as
\begin{eqnarray}\label{pressurefit2}
L(p)=\kappa_1p\exp\left(-\kappa_2p-\kappa_3p^2\right),
\end{eqnarray}
with the constants $\kappa_1,\kappa_2,\kappa_3$, which are obtained from the fit shown in Fig.~\ref{2DMOTFIG3}. At the curve's maximum, the fit yields $\kappa_2p/\kappa_3p^2=8$, showing that the collisions which limit the $^{40}$K-MOT capture rate are mainly the collisions with the hot background atoms, consisting mostly of $^{39}$K.

%
%
The background atoms are predominantly potassium atoms. These can collide either with the excited or the non-excited $^{40}$K-atoms of the atomic beam. Depending on the isotopes of the colliding partners, these collisions have different cross sections. Collisions between an excited and a non-excited atom of the same isotope usually have a very large cross section due to the strong resonant dipole-dipole interaction, described by a $C_3/R^3$-potential. In 2D-MOT systems of other atomic species these collisions have been identified as the ones which limit the flux of the 2D-MOT~\nocite{DieSpr98}\cite{DieSpr98,SchBat02,ChaRoy06}. In the case of $^{40}$K, the scattering rate for these collisions is reduced by the small abundance of $^{40}$K in the vapor. Therefore other collisions might limit the flux. In order to identify the flux-limiting collisions we calculate the cross section of different possible collisions and deduce the corresponding collision rates. The cross sections can be calculated using the approach described in ref.~\cite{SteCho92} for losses out of a cold atom cloud. The cross section for collisions involving an excited and a non-excited $^{40}$K-atom is given by~\cite{SteCho92} 
\begin{eqnarray}\label{sigma4040}
\sigma_\textrm{\scriptsize eff}^{40,40^*}=\pi\left(\frac{4C_3}{mv_\textrm{\scriptsize esc}\left\langle v\right\rangle}\right)^{2/3},
\end{eqnarray}
where $m$ is the mass of the $^{40}$K-atom, $v_\textrm{\scriptsize esc}\sim1$\,m/s is the estimated transverse velocity kick needed to make an atom miss the $^{40}$K-MOT, and $C_3=5.4\times10^{-48}$\,Jm$^3$ is the dispersion coefficient for the resonant dipole-dipole interaction~\cite{DerJoh02}. The cross section for collisions involving a non-excited $^{40}$K-atom and a non-excited K-atom of the different isotopes is given by~\cite{SteCho92}
\begin{eqnarray}\label{sigma4039}
\sigma_\textrm{\scriptsize eff}^{40,39}\sim\sigma_\textrm{\scriptsize eff}^{40,41}\sim\sigma_\textrm{\scriptsize eff}^{40,40}=\pi\left(\frac{15\pi C_6}{8mv_\textrm{\scriptsize esc}\left\langle v\right\rangle}\right)^{1/3},
\end{eqnarray}
where $C_6=3.7\times10^{-76}$\,Jm$^6$ is the dispersion coefficient for the underlying van der Waals interaction~\cite{DerJoh02}. Substituting the experimental parameters, one obtains: $\sigma_\textrm{\scriptsize eff}^{40,40^*}=2.7\times10^{-16}\,\textrm{m}^2$ and $\sigma_\textrm{\scriptsize eff}^{40,39}\sim\sigma_\textrm{\scriptsize eff}^{40,41}\sim\sigma_\textrm{\scriptsize eff}^{40,40}=1.3\times10^{-17}\,\textrm{m}^2$. The resulting collision rates are proportional to the atomic densities $n_{39},\ n_{40}$ and $n_{41}$ of the corresponding isotopes in the vapor and the relative number of excited $^{40}$K-atoms in the atomic beam, which was estimated to $P\sim0.1$ for the given beam detunings and intensities. One obtains
\begin{eqnarray}\label{2DMOTcollisionrates}
\Gamma_\textrm{\scriptsize coll}^{40,40^*}\!\!&=&Pn_{40}\sigma_\textrm{\scriptsize eff}^{40,40^*}\left\langle v\right\rangle\ =\ 4.4\times10^{-16}n_{\textrm{\scriptsize K}},\\ \label{colldom}
\Gamma_\textrm{\scriptsize coll}^{40,39}&=&(1-P)n_{39}\sigma_\textrm{\scriptsize eff}^{40,39}\left\langle v\right\rangle \ =\ 4.4\times10^{-15}n_{\textrm{\scriptsize K}},\\
\Gamma_\textrm{\scriptsize coll}^{40,40}&=&(1-P)n_{40}\sigma_\textrm{\scriptsize eff}^{40,40}\left\langle v\right\rangle\ =\ 2.0\times10^{-16}n_{\textrm{\scriptsize K}},\\
\Gamma_\textrm{\scriptsize coll}^{40,41}&=&(1-P)n_{41}\sigma_\textrm{\scriptsize eff}^{40,41}\left\langle v\right\rangle\ =\ 3.0\times10^{-16}n_{\textrm{\scriptsize K}}
\end{eqnarray}
($n_\textrm{\scriptsize K}$ denoting the atomic density of potassium in the vapor cell). The dominant collision rate here is $\Gamma_\textrm{\scriptsize coll}^{40,39}$ (Eq.~(\ref{colldom})) for collisions involving a non-excited $^{40}$K-atom and a non-excited $^{39}$K-atom from the background. The largest collision rate for collisions between two $^{40}$K-atoms, $\Gamma_\textrm{\scriptsize coll}^{40,40^*}$, is by a factor of 10 smaller than $\Gamma_\textrm{\scriptsize coll}^{40,39}$. Therefore, collisions involving two $^{40}$K-atoms are not the collisions which limit the flux of the 2D-MOT. This is in contrast to 2D-MOT systems of other species. From the difference between $\Gamma_\textrm{\scriptsize coll}^{40,40^*}$ and $\Gamma_\textrm{\scriptsize coll}^{40,39}$ we conclude that the flux of the 2D-MOT for $^{40}$K could still be improved by about a factor of 10 by using a potassium sample of a higher isotopic enrichment.

\section{$^{6}$Li-$^{40}$K dual-species MOT}\label{DoubleMOT}

\paragraph{Introduction}
Previously, several groups have studied samples of two atomic species in a magneto-optical trap~\nocite{SanNus95,SchEng99,TelGar01,GolPap02,TagVoi06,SpiTre10}\cite{SanNus95,SchEng99,TelGar01,GolPap02,TagVoi06,SpiTre10,Tie09}. Here we report on the implementation and performance of our $^{6}$Li-$^{40}$K dual-species MOT and on the study of collisions between atoms of the different species. After a description of the experimental setup, we start with a characterization of the single-species MOTs and then focus on the collisions in the dual-species MOT.

\paragraph{Principle of operation} In a magneto-optical trap six \linebreak counter-propagating red-detuned overlapping laser beams cool and magneto-optically confine atoms in a magnetic quadrupole field around its zero~\cite{MetStr99}. MOTs for alkali-atoms require laser light of two frequencies, namely the cooling and the repumping frequency. The latter ensures that the atoms stay in the cycling transition used for cooling. Typically the repumping light has a much lower power than the cooling light as the atoms principally occupy the states belonging to the cooling transition. For $^{6}$Li, however, the power of the repumping light needs to be relatively high, since $^{6}$Li has a very small hyperfine structure in the excited state manifold (of the order of the linewidth). When laser cooled, $^{6}$Li-atoms thus very likely quit the cooling transition. Therefore, the repumping light needs to contribute to the cooling process. As a consequence it needs to be present in all six directions with the same polarization as the cooling light. Therefore, we use bichromatic MOT-beams containing both cooling and repumping frequencies. We adapt the same strategy also for $^{40}$K. 

\paragraph{Experimental setup}
Light for the dual-species MOT is derived from two bichromatic laser beams, containing each a cooling and a repumping frequency, which are provided by two separate optical fibers originating from the respective laser systems. The beams are superimposed using a dichroic mirror and then expanded by a telescope to a 1/e$^2$-diameter of 22\,mm. All subsequent beam reflections are realized by two-inch sized broadband mirrors (Thorlabs, ref.~BB2-E02-10). The beam is separated by three two-inch sized broadband polarization cubes (Lambda Optics, ref.~BPB-50.8SF2-550) into four arms that form a partially retro-reflected MOT, in which only the vertical beam pair is composed of independent counter-propagating beams. Each retro-reflected MOT beam is focused with a lens of focal length 10\,cm, placed at a distance of $\sim11\,$cm in front of the retro-reflecting mirror, in order to increase the intensity and therefore compensate for the losses in the optics and the light absorption by the trapped atoms. The distribution of the light power over the MOT beams is independently adjusted for the two wavelengths using a pair of custom-made wave plates, placed in front of each broad-band splitting cube. The wave plate pair consists of a $\lambda/2$ plate of order 4 for the wavelength 767\,nm and a $\lambda/2$ plate of order 4 for the wavelength 671\,nm. To a very good approximation each of these wave plates can turn the polarization direction for one wavelength without affecting the polarization for the other one (since it is $4.5\times767\approx 5\times 671$ and $4.5\times671\approx 4\times 767$). The circular polarization of the MOT beams is produced by first order $\lambda/4$ plates for 767\,nm, which work sufficiently well also for 671\,nm. All four frequency components thus have the same circular polarizations in each beam. A mechanical shutter is placed in the focus of the telescope allowing to produce total extinction of the MOT light in addition to the partial and fast switching by the AOMs.

The bichromatic beam for the $^{40}$K-MOT has a total power of $P_\textrm{\scriptsize fiber}=220$\,mW and its frequencies are red-detuned by $\sim3\,\Gamma$ from the $4S_{1/2}(F=9/2)\rightarrow4P_{3/2}(F'=11/2)$ cooling and by $\sim5\,\Gamma$ from the $4S_{1/2}(F=7/2)\rightarrow4P_{3/2}(F'=9/2)$ repumping transition (see Fig.~\ref{Levels}). The intensity of the cooling light is $\sim20$ times bigger than that of the repumping light. The bichromatic beam for the $^{6}$Li-MOT has a total power of $P_\textrm{\scriptsize fiber}=110$\,mW and its frequencies are red-detuned by $\sim5\,\Gamma$ from the $2S_{1/2}(F=3/2)\rightarrow2P_{3/2}(F'=5/2)$ cooling and by $\sim3\,\Gamma$ from the $2S_{1/2}(F=1/2)\rightarrow2P_{3/2}(F'=3/2)$ repumping transition (Fig.~\ref{Levels}). The power of the cooling light is $\sim5$ times bigger than that of the repumping light.

The magnetic field for the dual-species MOT is created by a pair of coils in anti-Helmholtz configuration. The magnetic field gradient along the vertically directed symmetry axis is $\partial_z B=8\,$G/cm. This gradient yields an optimum atom number for the $^{40}$K-MOT.

The atoms in the dual-species MOT are probed by absorption imaging. In order to obtain a two-dimensional density profile of the atom cloud, three pictures are taken and recorded by a CCD-camera (PCO imaging, ref.~Pixelfly qe). The first picture is taken with the imaging beam tuned near resonance and thus records the shadow cast by the atom cloud on the CCD-chip of the camera. The second picture is taken with the imaging beam tuned far off resonance (by $-10\,\Gamma$) and records the intensity profile of the imaging beam. The third picture is taken in absence of the imaging beam and records the background signal. The change of frequency of the imaging beam allows to take the first two pictures with a short time delay (2\,ms), while keeping the imaging beam at the same frequency would require to wait for the atom cloud to disappear before the second picture could be recorded. Thus, the intensity fluctuations of the imaging beam during the recording process are minimized and both pictures can be taken with the same intensity. 

Each atomic species requires its own imaging beam, which is provided by a separate optical fiber originating from the respective laser system (see Fig.~\ref{Optics}). The two imaging beams are superimposed using a dichroic mirror and expanded by a telescope to a 1/e$^2$-diameter of $27.5$\,mm. The imaging beams have low intensity ($I_\textrm{\scriptsize img}\sim0.01I_\textrm{\scriptsize sat}$ in the beam center), are circularly polarized and pass through the MOT along the horizontal direction, perpendicular to the axis of the quadrupole magnetic field of the MOT. No bias magnetic field is applied when absorption pictures are taken. The best atom number estimate from the measured absorption pictures is thus given by using an averaged squared Clebsch-Gordan coefficient, which is $C^2=0.5$ for $^{6}$Li and $C^2=0.4$ for $^{40}$K. Both beams are red detuned by $2\,\Gamma$ from the $4S_{1/2}(F=9/2)\rightarrow4P_{3/2}(F'=11/2)$ and the $2S_{1/2}(F=3/2)\rightarrow2P_{3/2}(F'=5/2)$ cooling transitions of $^{40}$K and $^{6}$Li, respectively (see Fig.~\ref{Levels}), so as to reduce saturation effects. For the chosen length of the imaging pulses (100\,$\mu$s) no repumping is required during the imaging process (we verified for $^{6}$Li that even in the case of a resonant imaging beam, the presence of a repumping beam would yield an increase of the detected atom number of only $8\%$ , which would be even less for $^{40}$K). In order to image the total number of atoms in the MOTs the atom clouds are exposed for 500\,$\mu s$ to only the repumping light before the image is taken in order to optically pump all atoms to the hyperfine ground state which is imaged. The overall uncertainty of the absolute atom number determination is estimated to be $50$\%. 

\paragraph{Experimental results}
In single-species operation we characterized the MOTs using the parameters for the optimized dual-species operation. We determined the atom numbers, the atomic densities in the cloud center, the loading times and the temperatures. Furthermore, we studied for each atomic species the dependence of the steady-state MOT atom number on the following parameters: the power and detuning of the cooling light and the intensity ratio between the repumping and cooling light. In dual-species operation, we studied the dependence of heteronuclear light-induced cold collisions on the laser power used for the MOT-beams. The optimum parameters, which lead to atom numbers of $N_\textrm{\scriptsize single}\sim8.9\times10^9$ in the $^{40}$K-MOT and $N_\textrm{\scriptsize single}\sim5.4\times10^9$ in the $^{6}$Li-MOT, are displayed in Tab.~\ref{MOTparameters} together with the characteristics of the MOTs (in dual-species operation, the atom numbers only slightly change due to the additional interspecies collisions to \linebreak $N_\textrm{\scriptsize dual}\sim8.0\times10^9$ in the $^{40}$K-MOT and $N_\textrm{\scriptsize dual}\sim5.2\times10^9$ in the $^{6}$Li-MOT). The $(1-1/e)$-loading times of the MOTs are $\sim5$\,s for $^{40}$K and $\sim6$\,s for $^{6}$Li.
%
\begin{table}[h]
	\centering
		\begin{tabular}{p{3cm} c c} 
		  & $^{40}$K-MOT\hspace{1cm} & $^{6}$Li-MOT\\ \hline\hline
			$P_\textrm{\scriptsize fiber}\ [\textrm{mW}]$ & 220 & 110\\
			$\Delta\omega_\textrm{\scriptsize cool}\ [\Gamma]$ & -3 & -5\\
			$\Delta\omega_\textrm{\scriptsize rep}\ [\Gamma]$ & -5 & -3\\
			$\Gamma/(2\pi)\ [\textrm{MHz}]$ & 6.04 & 5.87 \\
			$I_\textrm{\scriptsize cool}$ per beam $[I_\textrm{\scriptsize sat}]$ & 13 & 4\\
			$I_\textrm{\scriptsize sat}\ [\textrm{mW/cm}^2]$ & 1.75 & 2.54\\
			$I_\textrm{\scriptsize rep}/I_\textrm{\scriptsize cool}$ & 1/20 & 1/5\\
			$\partial_z B\ [\textrm{G/cm}]$ & 8 & 8\\
			$N_\textrm{\scriptsize single}\ [\times10^9]$ & 8.9 & 5.4\\
			$N_\textrm{\scriptsize dual}\ [\times10^9]$ & 8.0 & 5.2\\ 
			$n_\textrm{\scriptsize c}\ [\times10^{10}\,\textrm{at./cm}^3]$ & 3 & 2\\
			$T\ [\mu\textrm{K}]$ & 290 & 1400\\ \hline\hline
		\end{tabular}
	\caption[3cm]{Characteristic parameters of the dual-species $^{6}$Li-$^{40}$K-MOT.}
	\label{MOTparameters}
\end{table}
%

Magneto-optical traps with large atom numbers have a high optical density and are optically dense for weak resonant laser beams. Therefore, when determining the atom number via absorption imaging, the frequency of the imaging beam has to be detuned, so not to ``black out'' the image.

\begin{figure}[h]
\centering
\includegraphics[width=8.7cm]{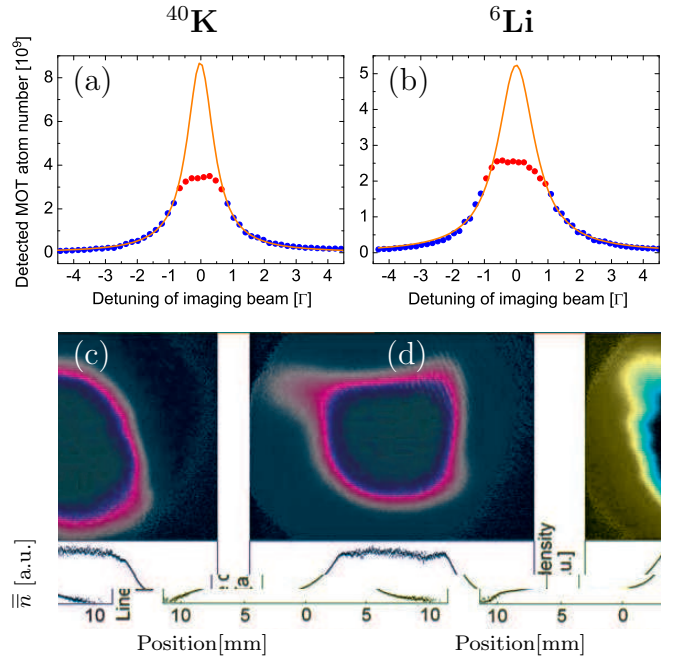}
\caption{(Color online) (a,b) Detected atom number in the MOTs as a function of the detuning of the imaging beams. Circles correspond to the experimental data and solid curves to Lorentzian fits of the curve wings with an imposed width, which was determined by another measurement. (c,d) Absorption images of the MOTs and the doubly-integrated optical density profile $\overline{\overline{n}}$, recorded with a resonant imaging beam. The graphs (a,c) relate to the $^{40}$K-MOT and (b,d) to the $^{6}$Li-MOT. The flat top of $\overline{\overline{n}}$ in the graphs (c,d) and the saturation of the detected atom number for small magnitudes of the detuning in the graphs (a,b) demonstrate that the MOTs are optically dense for the imaging beam when the detuning is small. Their (extrapolated) central optical densities for a resonant imaging beam are $\sim20$ for $^{40}$K and $\sim15$ for $^{6}$Li.}\label{LiKMOTFIG1}
\end{figure}

Figures~\ref{LiKMOTFIG1} (a,b) depict the \textit{detected} atom number of the two MOTs (circles) as a function of the detuning of the imaging beam. The detected atom number was derived from the measured optical density assuming the imaging beam to be resonant. The curves are expected to have the shape of a Lorentzian with the peak centered around zero detuning. The experimental data shown in Figures~\ref{LiKMOTFIG1} (a,b) clearly deviate from a Lorentzian behavior---they saturate for small magnitudes of the detuning. This deviation demonstrates that the MOTs are optically dense for small detunings. A correct estimate of the atom number is obtained from an extrapolation of the experimental data to zero detuning based on a Lorentzian fit of the curve wings (solid curves). A reliable extrapolation, however, requires imposing the width of the Lorentzian fit. In order to determine this width, an additional experiment was done (not shown): the data in Figs.~\ref{LiKMOTFIG1} (a,b) were again recorded and fitted by a Lorentzian for a MOT with a small atom number and a low optical density (obtained by a short loading of 250\,ms). The widths found by this additional measurement were $1.05\,\Gamma$ for $^{40}$K and $1.5\,\Gamma$ for $^{6}$Li. For $^{40}$K this width corresponds to the natural linewidth of the exited state addressed by the imaging transition. For $^{6}$Li the width is larger than the natural linewidth, since the small excited hyperfine structure is unresolved and thus its width ($\sim0.5\,\Gamma$) and the natural linewidth add up (this line broadening does not occur when a bias magnetic field is applied and a closed transition is used for imaging). The peak values of the Lorentzian fits in Figs.~\ref{LiKMOTFIG1} (a,b) finally yield the atom numbers in the MOTs, given in Tab.~(\ref{MOTparameters}).

%
%

Figures~\ref{LiKMOTFIG1} (c,d) show images of the MOTs and their doubly-integrated optical density profiles $\overline{\overline{n}}$ for the case of a resonant imaging beam. The flat top of $\overline{\overline{n}}$ as a function of position shows that the MOTs are optically dense. Their central optical densities for the resonant imaging beam are determined to be $\sim20$ for $^{40}$K and $\sim15$ for $^{6}$Li by the extrapolation technique described above. In addition, the density profiles in Figs.~\ref{LiKMOTFIG1} (c,d) show that the MOTs have spatial extensions of the order of $1$\,cm.

The atomic density in the MOT center is extracted from the recorded two-dimensional density profile as follows. The recorded profile is proportional to the atomic density $n(x,y,z)$ integrated along the imaging beam direction $z$: $g(x,y)\propto\int n(x,y,z)dz$. When assuming that the MOT has cylindrical symmetry (with the symmetry axis along the $x$-direction), the local atomic density $n_\textrm{\scriptsize c}$ at the MOT center is given by the maximum of the inverse Abel transform of $g(x_\textrm{\scriptsize c},y)$, where $x_\textrm{\scriptsize c}$ is the $x$-coordinate of the MOT center
\begin{eqnarray}\label{Abel}
    n_\textrm{\scriptsize c}=\max\limits_{r}\left(-\frac{1}{\pi}\int_r^\infty\left(\frac{\partial g(x_\textrm{\scriptsize c},y)}{\partial y}\right)\frac{\textrm{d}y}{\sqrt{y^2-r^2}}\right),
\end{eqnarray}
with $r=\sqrt{y^2+z^2}$ denoting the distance to the MOT center~\cite{DriOss02}. Since the derivative $\partial g/\partial y$ is very sensitive to noise, the density profile $g$ is smoothened before its derivative is calculated. We obtain $n_\textrm{\scriptsize c}^\textrm{\scriptsize K}\sim3\times10^{10}$\,atoms/cm$^3$ and $n_\textrm{\scriptsize c}^\textrm{\scriptsize Li}\sim2\times10^{10}$\,atoms/cm$^3$, respectively.

The temperature of the MOTs in single-species operation was determined by the time-of-flight method~\cite{MetStr99}. The $^{40}$K-MOT has a temperature of 290\,$\mu$K and the $^{6}$Li-MOT of 1.4\,mK. Both temperatures are higher than the Doppler cooling limit, because of the high intensity in the MOT beams. In addition, for $^{6}$Li, the unresolved excited hyperfine structure (see Fig.~\ref{Levels}) inhibits sub-Doppler cooling effects. The same temperatures are found in dual-species operation. The measured temperatures and atomic densities yield the peak phase space densities $D_\textrm{\scriptsize K}=n_\textrm{\scriptsize c}^\textrm{\scriptsize K}\Lambda_\textrm{\scriptsize K}^3\sim1.2\times10^{-7}$ and $D_\textrm{\scriptsize Li}=n_\textrm{\scriptsize c}^\textrm{\scriptsize Li}\Lambda_\textrm{\scriptsize Li}^3\sim1.3\times10^{-7}$ with the thermal de Broglie wavelength $\Lambda=\sqrt{2\pi\hbar^2/(mk_\textrm{\scriptsize B}T)}$, respectively.

\begin{figure}[h]
\centering
\includegraphics[width=8.7cm]{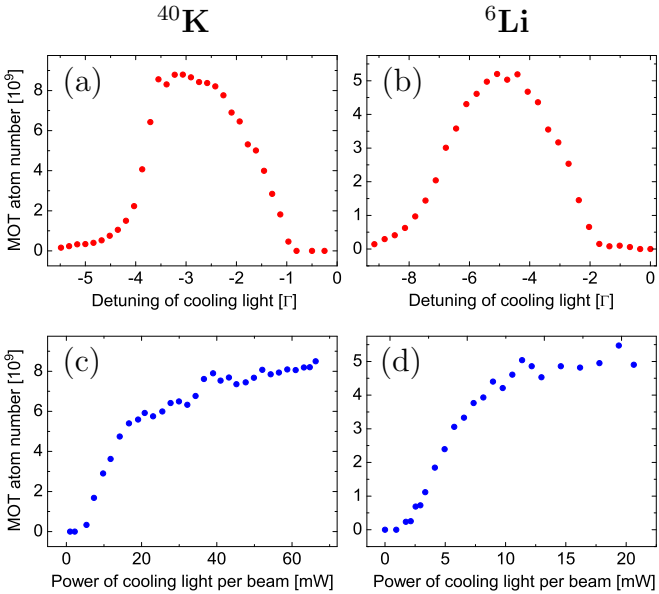}
\caption{(Color online) MOT atom number as a function of (a,b) the detuning and (c,d) the power of the cooling light per MOT beam for a constant intensity ratio between the cooling and repumping light. The graphs (a,c) relate to the $^{40}$K-MOT and (b,d) to the $^{6}$Li-MOT. For $^{40}$K a power of 45\,mW corresponds to an intensity of $13\,I_\textrm{\scriptsize sat}$, for $^{6}$Li a power of 20\,mW corresponds to an intensity of $4\,I_\textrm{\scriptsize sat}$, with the respective saturation intensities $I_\textrm{\scriptsize sat}$ given in Tab.~\ref{MOTparameters}.}\label{LiKMOTFIG2}
\end{figure}

The dependence of the MOT atom number on the detuning of the cooling light is depicted in Figs.~\ref{LiKMOTFIG2} (a,b). The atom number is maximum at $\Delta\omega_\textrm{\scriptsize cool}^\textrm{\scriptsize K}=-3\,\Gamma$ for $^{40}$K and at $\Delta\omega_\textrm{\scriptsize cool}^\textrm{\scriptsize Li}=-5\,\Gamma$ for $^{6}$Li, and has a FWHM of $2.3\,\Gamma$ and $4.1\,\Gamma$, respectively. 
%
%

Figures~\ref{LiKMOTFIG2} (c,d) show the dependence of the MOT atom number on the power of the cooling light per MOT beam. In the figures, a power of 10\,mW corresponds to an on-resonance peak intensity of $\sim3\,I_\textrm{\scriptsize sat}$ (Fig.~\ref{LiKMOTFIG2} (c)) and $\sim2\,I_\textrm{\scriptsize sat}$ (Fig.~\ref{LiKMOTFIG2} (d)) in each of the six MOT beams. The atom number increases with increasing light power and saturates for higher powers. The saturation is due to several effects. First, the absorption probability for the cooling light saturates for high intensities. Second, the repulsive forces between the atoms due to rescattered photons and the temperature of the cloud increase with increasing light power~\cite{SteCho92}. Finally the scattering rate for light-induced cold collisions increases with increasing light power.

%
\begin{figure}[h]
\centering
\includegraphics[width=8.7cm]{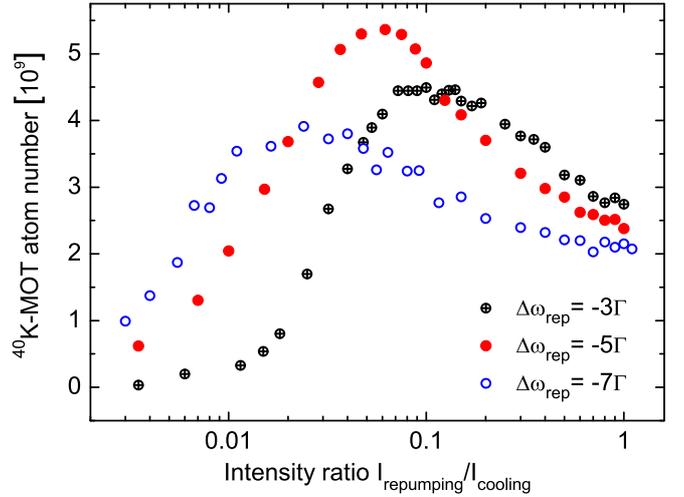}
\caption{(Color online) $^{40}$K-MOT atom number as a function of the intensity ratio between repumping and cooling light for three different repumping detunings $\Delta\omega_\textrm{\scriptsize rep}$ and a constant cooling light power of 18\,mW per MOT beam (which corresponds to an intensity of $6$\,$I_\textrm{\scriptsize sat}$).}\label{KMOTFIG1}
\end{figure}
%

\begin{figure}[h]
\centering
\includegraphics[width=8.7cm]{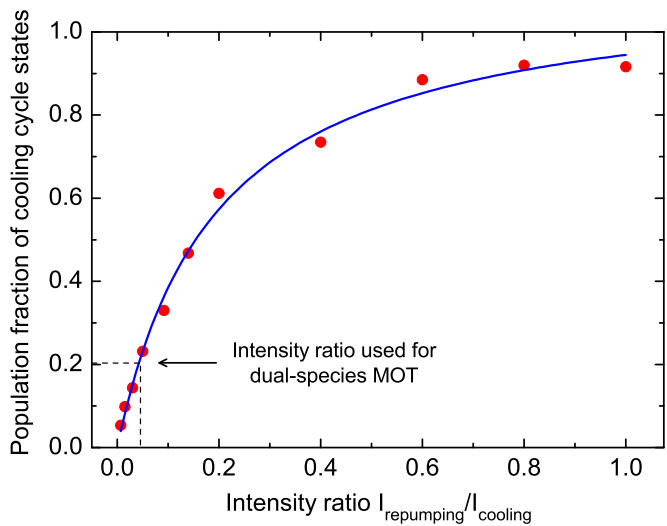}
\caption{(Color online) Circles: measured fraction of atoms in the $^{40}$K-MOT populating the states $F=9/2$ or $F'=11/2$ (cooling cycle states) as a function of the intensity ratio between repumping and cooling light for the repumping detuning $\Delta\omega_\textrm{\scriptsize rep}=-5\,\Gamma$ and a constant cooling light power of 18\,mW per MOT beam (which corresponds to an intensity of $6$\,$I_\textrm{\scriptsize sat}$). For the ratio which maximizes the total atom number in the $^{40}$K-MOT, $I_\textrm{\scriptsize rep}/I_\textrm{\scriptsize cool}\sim1/20$, only 20\% of the trapped atoms occupy the cooling cycle states. Solid curve: a fit based on Einstein's rate equations.}\label{KMOTFIG2}
\end{figure}

Figure~\ref{KMOTFIG1} shows the dependence of the $^{40}$K-MOT atom number on the intensity ratio $I_\textrm{\scriptsize rep}/I_\textrm{\scriptsize cool}$ between repumping and cooling light for three different repumping detunings $\Delta\omega_\textrm{\scriptsize rep}^{(1)}=-3\,\Gamma$, $\Delta\omega_\textrm{\scriptsize rep}^{(2)}=-5\,\Gamma$ and $\Delta\omega_\textrm{\scriptsize rep}^{(3)}=-7\,\Gamma$ and a constant cooling light power of 18\,mW per MOT beam. The curves have a maximum at different ratios $I_\textrm{\scriptsize rep}/I_\textrm{\scriptsize cool}$, the position of the maxima lying at higher ratios for lower detunings. Furthermore, the maxima have different values for the three curves. The maximum is biggest for the detuning $\Delta\omega_\textrm{\scriptsize rep}^{(2)}=-5\,\Gamma$. The shape of the curves can be understood as follows. Each curve increases between $I_\textrm{\scriptsize rep}/I_\textrm{\scriptsize cool}=0$ and the position of the maximum, because the transition probability of the repumping transition increases with increasing repumping intensity. Thus the atoms are more efficiently cooled by the cooling light, as they are more efficiently repumped into the cycling transition. However, when the intensity of the repumping light becomes too large, the curve decreases again. Then, due to the strong repumping, the atoms are exposed to the more intense near-resonant cooling light, which causes light-induced cold collisions, leading to trap loss. At the maximum, the repumping is sufficiently strong to allow for an efficient cooling, and it is sufficiently weak to preserve the atoms from cold collisions induced by the strong cooling light. The value of the curve maximum is biggest for the detuning $\Delta\omega_\textrm{\scriptsize rep}^{(2)}=-5\,\Gamma$. It is situated at $I_\textrm{\scriptsize rep}/I_\textrm{\scriptsize cool}\sim1/20$, for which, as one can see below, only $\sim20\%$ of the $^{40}$K-MOT atoms occupy the cooling cycle states $F=9/2$ or $F'=11/2$ (see Fig.~\ref{KMOTFIG2}), the others occupying the ``dark'' hyperfine ground state $F=7/2$.

For very small intensity ratios $I_\textrm{\scriptsize rep}/I_\textrm{\scriptsize cool}\leq0.01$ the atom number in the $^{40}$K-MOT is larger for higher repumping detunings (Fig.~\ref{KMOTFIG1}). This behavior might be a consequence of the fact that the $^{40}$K-MOT is loaded from a slow atomic beam. The beam atoms, which have a negative Doppler shift of more than $5\,\Gamma$ with respect to the counter-propagating MOT beams, might absorb the repumping light more likely when it has a higher detuning. 

%
%

Figure~\ref{KMOTFIG2} shows the fraction of atoms in the $^{40}$K-MOT (circles) which populate the states $F=9/2$ or $F'=11/2$ (i.e.~the cooling cycle states, see Fig.~\ref{Levels}) as a function of the intensity ratio $I_\textrm{\scriptsize rep}/I_\textrm{\scriptsize cool}$ between repumping and cooling light. In the experiment, the cooling light power was fixed to 18\,mW per MOT beam, and the repumping detuning was $\Delta\omega_\textrm{\scriptsize rep}=-5\,\Gamma$. The graph was recorded as follows. The absolute population of the states $F=9/2$ and $F'=11/2$ was measured by simultaneously switching off both the repumping and cooling light of the $^{40}$K-MOT 600\,$\mu s$ before taking the image (with the imaging beam being near-resonant with the $F=9/2\rightarrow F'=11/2$-transition). During the 600\,$\mu s$ time delay, all excited atoms relax to one of the ground states. For the used detunings and intensities of the MOT-beams $\sim90\%$ of the excited atoms occupy the state $F'=11/2$ and thus relax to the ground state $F=9/2$, which is imaged. Therefore, the image approximately yields the sum of the populations of the states $F=9/2$ and $F'=11/2$. The total population of all states (i.e.~the total number of trapped atoms) was measured as described in the previous paragraph.

The curve in Fig.~\ref{KMOTFIG2} is increasing with increasing ratios $I_\textrm{\scriptsize rep}/I_\textrm{\scriptsize cool}$ and it saturates for high ratios. For the ratio $I_\textrm{\scriptsize rep}/I_\textrm{\scriptsize cool}=1/5$ about $60\%$ of the $^{40}$K-MOT atoms occupy the cooling cycle states. For this ratio the fluorescence emitted by the $^{40}$K-MOT is found to be maximum. For the ratio $I_\textrm{\scriptsize rep}/I_\textrm{\scriptsize cool}=1/20$, which is used in the experiment, only $\sim20\%$ of the atoms occupy the cooling cycle states. Atom losses due to light-induced collisions are thus minimized.

The solid curve in Fig.~\ref{KMOTFIG2} shows a fit of the experimental data, based on a simple model, assuming $^{40}$K to be a four-level atom (with the states $F=9/2$, $F=7/2$, $F'=11/2$ and $F'=9/2$). Einstein's rate equations yield that the curve obeys the law $P_\textrm{\scriptsize ccs}=1/(1+a+b/(I_\textrm{\scriptsize rep}/I_\textrm{\scriptsize cool}))$, with the fitting parameters $a=-0.11$ and $b=0.17$, which depend on the transition probabilities and the used intensities and detunings.

\begin{figure}[h]
\centering
\includegraphics[width=8.7cm]{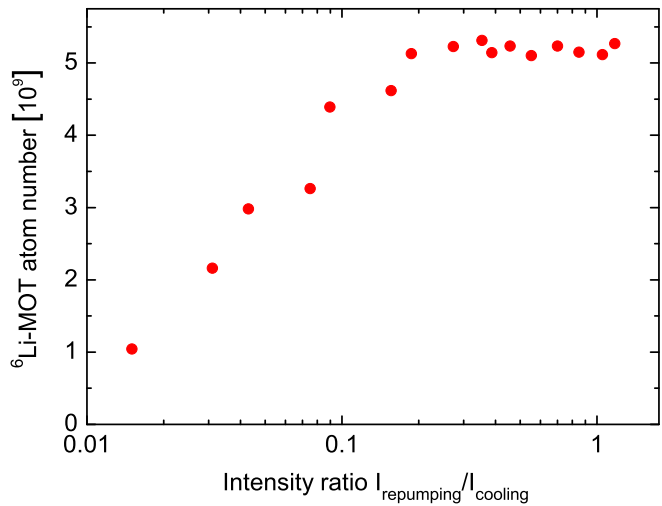}
\caption{(Color online) $^{6}$Li-MOT atom number as a function of the intensity ratio between repumping and cooling ligh for a constant cooling light power of 11\,mW per MOT beam (which corresponds to an intensity of $2$\,$I_\textrm{\scriptsize sat}$). In comparison to $^{40}$K (Fig.~\ref{KMOTFIG1}), the optimum atom number requires a larger intensity in the repumping light, which is a consequence of the unresolved excited hyperfine structure of $^{6}$Li.}\label{LiMOTFIG1}
\end{figure}

Figure~\ref{LiMOTFIG1} shows the dependence of the $^{6}$Li-MOT atom number on the intensity ratio $I_\textrm{\scriptsize rep}/I_\textrm{\scriptsize cool}$ between repumping and cooling light for the repumping detuning $\Delta\omega_\textrm{\scriptsize rep}=-3\,\Gamma$ and a constant cooling light power of 11\,mW per MOT beam. In contrast to Figure~\ref{KMOTFIG1}, the curve does not have a maximum but rather increases with increasing $I_\textrm{\scriptsize rep}/I_\textrm{\scriptsize cool}$ and saturates. This behavior is a result of the important contribution of the repumping light to the cooling process, particular to $^{6}$Li, as it has an unresolved excited state hyperfine structure.

%
%

In a dual-species MOT, inelastic collisions between atoms of the two different species can occur and represent important loss mechanisms. Previous studies have shown that the principal loss mechanisms for heteronuclear collisions in dual-species MOTs involve one ground-state and one excited atom of different species~\cite{SchEng99,TelGar01}. Such atom pairs can undergo radiative escape or fine-structure changing collisions~\cite{WeiZim03}. Both these loss processes require the two atoms to approach each other sufficiently close such that a large enough interaction energy is gained to make the atoms leave the trap. The long-range behavior of the scattering potentials determines if the atoms can approach each other sufficiently. For LiK, the scattering potentials for a singly-excited heteronuclear atom pair are all attractive for the case where the K atom is excited and all repulsive for the case where the Li atom is excited~\cite{BusAch87}. As a consequence, a ground-state K atom and an excited Li atom repel each other and are prevented from undergoing inelastic collisions (optical shielding). Inelastic collisions involving singly-excited heteronuclear atom pairs thus always contain an excited K atom. In order to minimize the rate of heteronuclear collisions in the LiK-MOT, the density of excited K atoms must therefore be reduced. Furthermore, the atomic density in the trap as well as the relative speed of the colliding atoms, i.e.~the temperature of the cloud, need to be minimized.

In our $^{6}$Li-$^{40}$K dual-species MOT the following strategy is applied in order to minimize inelastic heteronuclear collisions. First the use of very low magnetic field gradients (8\,G/cm), which decreases the atomic densities ($n_\textrm{\scriptsize c}^\textrm{\scriptsize K}\sim3\times10^{10}$\,atoms/cm$^3$ and $n_\textrm{\scriptsize c}^\textrm{\scriptsize Li}\sim2\times10^{10}$\,atoms/cm$^3$). Second, low intensities in the repumping light for both, $^{6}$Li and $^{40}$K, are used in order to decrease the number of excited atoms. Decreasing the number of excited $^{6}$Li atoms here a priori serves to decrease the temperature of the $^{6}$Li-cloud. Since that is much larger than the temperature of the $^{40}$K-cloud, the relative speed of two colliding atoms and thus the collision rate can be efficiently decreased by minimizing the temperature of the $^{6}$Li-cloud. Finally a small mutual influence of the MOTs is obtained: the atom numbers in the MOTs decrease by $\sim 4$\% in the $^{6}$Li-MOT and $\sim 10$\% in the $^{40}$K-MOT due to the presence of the other species.

\begin{figure}[h]
\centering
\includegraphics[width=8.7cm]{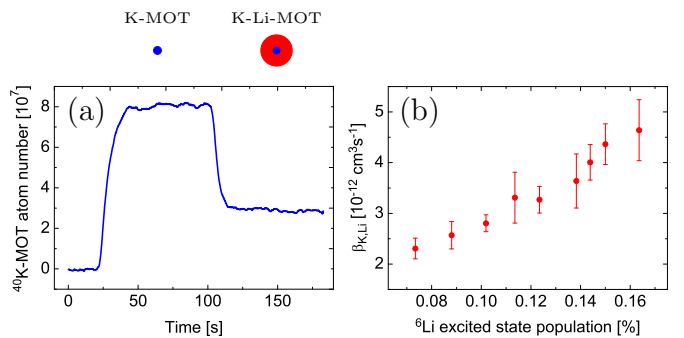}
\caption{(Color online) (a) Evolution of the atom number in the $^{40}$K-MOT in the absence ($t<100$\,s) and presence ($t>100$\,s) of the $^{6}$Li-MOT for an increased magnetic field gradient of 16\,G/cm. (b) Trap loss coefficient $\beta_\textrm{\scriptsize KLi}$ for heteronuclear collisions as a function of the relative excited state population of the trapped $^{6}$Li atoms.}\label{LiKMOTFIG3}
\end{figure}

The importance of decreasing the magnetic field gradients in order to minimize the heteronuclear collision rate in the dual-species MOT is demonstrated in Fig.~\ref{LiKMOTFIG3} (a), which depicts the effect of the $^{6}$Li-MOT on the $^{40}$K-MOT atom number when a two-times larger magnetic field gradient (16\,G/cm) is used. At this gradient the atomic density in the $^{6}$Li-MOT is by a factor of 4 larger than at the gradient used for the optimized MOT. In the experiment, the $^{40}$K-MOT was intentionally reduced in size (by decreasing the 2D-MOT flux) to ensure a better inclosure in the $^{6}$Li-MOT. The curve shows that $\sim65$\% of the $^{40}$K-MOT atoms leave the trap due to the enhanced heteronuclear collisions. Using a low magnetic field gradient is therefore helping significantly to decrease the heteronuclear collisions.

In the following we determine the trap loss coefficients for the (optimized) dual-species MOT in order to quantify the heteronuclear collisions. The rate equation for the atom number in a dual-species MOT (with species $A$ and $B$) reads~\cite{SchEng99}
\begin{eqnarray}\label{RateEQ}
\frac{\textrm{d}N_\textrm{\scriptsize A}}{\textrm{d}t}=L_\textrm{\scriptsize A}-\gamma N_\textrm{\scriptsize A}-\beta_\textrm{\scriptsize AA}\!\!\int n_\textrm{\scriptsize A}^2\textrm{d}V-\beta_\textrm{\scriptsize AB}\!\!\int n_\textrm{\scriptsize A}n_\textrm{\scriptsize B}\textrm{d}V,\quad
\end{eqnarray}
where $L_\textrm{\scriptsize A}$ is the loading rate, $\gamma$ the trap loss rate due to collisions with background gas atoms and $n_\textrm{\scriptsize A},n_\textrm{\scriptsize B}$ the local atomic densities. $\beta_\textrm{\scriptsize AA}$ and $\beta_\textrm{\scriptsize AB}$ denote the cold collision trap loss coefficients for homo- and heteronuclear collisions, respectively. $L_\textrm{\scriptsize A}$ and $\gamma$ are determined from the loading and decay curves of the single-species MOTs. The obtained values for $L_\textrm{\scriptsize A}$ are given in Tab.~\ref{MOTparameters} and $\gamma$ is found to be $1/7.5$\,s$^{-1}$. The homonuclear trap loss coefficients $\beta_\textrm{\scriptsize AA}$ are determined from the steady state atom numbers in single-species operation using the measured density profiles. For the experimental conditions indicated in Tab.~(\ref{MOTparameters}), we obtain
\begin{eqnarray}\label{traplosscoefficients}\label{betaLiLi}
\beta_\textrm{\scriptsize LiLi}&=&(8\pm 4)\times10^{-12}\,\textrm{cm}^{3}\textrm{s}^{-1},\\ \label{betaKK}
\beta_\textrm{\scriptsize KK}&=&(6\pm 3)\times10^{-13}\,\textrm{cm}^{3}\textrm{s}^{-1}.
\end{eqnarray}
The determination of the heteronuclear trap loss coefficients $\beta_\textrm{\scriptsize AB}$ for the optimized dual-species configuration would require the knowledge of the mutual overlap of the MOTs, which is difficult to estimate when absorption images are taken only along one direction. We therefore choose a configuration, which makes the determination of $\beta_\textrm{\scriptsize AB}$ less dependent on assumptions about the mutual overlap (but which does not change the value of $\beta_\textrm{\scriptsize AB}$). We reduce the atom flux of species $A$, in order to decrease the spatial extension of the trapped cloud of species $A$ and to place it in the center of the cloud of species $B$. A video camera which records the fluorescence of the MOTs from a different direction than that of the absorption imaging verifies that this configuration is indeed achieved. Then, in Eq.~(\ref{RateEQ}) it is $\int n_\textrm{\scriptsize A}n_\textrm{\scriptsize B}\textrm{d}V\sim n_\textrm{\scriptsize c}^\textrm{\scriptsize B} N_\textrm{\scriptsize A}$. Comparing the steady-state atom numbers for the different configurations then yields
\begin{eqnarray}\label{traplosscoefficients}\label{betaLiK}
\beta_\textrm{\scriptsize LiK}&=&(1\pm 0.5)\times10^{-12}\,\textrm{cm}^{3}\textrm{s}^{-1},\\ \label{betaKLi}
\beta_\textrm{\scriptsize KLi}&=&(3\pm 1.5)\times10^{-12}\,\textrm{cm}^{3}\textrm{s}^{-1},
\end{eqnarray}
for the experimental conditions indicated in Tab.~(\ref{MOTparameters}). Comparing all four trap loss coefficients, the dominant is $\beta_\textrm{\scriptsize LiLi}$ (Eq.~(\ref{betaLiLi})) for light-induced homonuclear $^{6}$Li-$^{6}$Li collisions. This is a consequence of the large temperature of the $^{6}$Li-MOT and the unresolved hyperfine structure of $^{6}$Li which prohibits the creation of a dark MOT, leading to a large excited state population. The much smaller homonuclear trap loss coefficient $\beta_\textrm{\scriptsize KK}$ for $^{40}$K (Eq.~(\ref{betaKK})) is consistent with Fig.~\ref{KMOTFIG1} which shows that, for $^{40}$K, small repumping intensities are favorable. The heteronuclear trap loss coefficients $\beta_\textrm{\scriptsize LiK},\beta_\textrm{\scriptsize KLi}$ (Eqs.~(\ref{betaLiK}) and (\ref{betaKLi})) are also much smaller than $\beta_\textrm{\scriptsize LiLi}$, indicating that our applied strategy for decreasing the heteronuclear collisions is good. In the Amsterdam group the heteronuclear trap loss coeffiecients were found by a factor of about 2 larger than ours~\cite{Tie09}. A dark SPOT MOT has been implemented in order to reduce the excited state population of the $^{40}$K atoms. In the next paragraph we show, however, that it is also important to reduce the excited state population of the $^{6}$Li atoms.
%
%

Figure~\ref{LiKMOTFIG3} (b) depicts the dependence of the trap loss coefficient $\beta_\textrm{\scriptsize KLi}$ on the relative excited state population of the $^{6}$Li atoms. The graph was obtained by recording the influence of the $^{6}$Li-MOT on the $^{40}$K-MOT as the power of the $^{6}$Li-MOT beams was varied. For each power it was verified that the $^{40}$K-MOT was placed in the center of the $^{6}$Li-MOT and the atomic density of the $^{6}$Li-MOT was recorded. In the experiment a magnetic field gradient of 16\,G/cm was used. The central atomic density of the $^{6}$Li-MOT was found to be approximately constant, when the power was varied ($n_\textrm{\scriptsize c}^\textrm{\scriptsize Li}\sim 8\times10^{10}$\,atoms/cm$^3$). The relative excited state population for a given beam power was estimated using Einstein's rate equations. In addition the variation of the excited state population was measured by recording the fluorescence emitted by the $^{6}$Li-MOT and by measuring the number of captured atoms. The latter changed by a factor of 1.5 in the considered range of beam powers. The graph in Fig.~\ref{LiKMOTFIG3} (b) shows that the trap loss coefficient increases by more than a factor of 2 as the relative excited state population is increased from $\sim 7$\% to $\sim 16$\%.  The error bars shown in the figure refer to statistical errors. The uncertainty due to systematic errors is estimated to be $50$\%. The significant increase of $\beta_\textrm{\scriptsize KLi}$ demonstrates the importance of minimizing the number of excited $^{6}$Li atoms (and not only that of the excited $^{40}$K atoms). One reason for this increase is the increase of temperature of the $^{6}$Li-MOT, which changes from $\sim1\,$mK to $\sim1.6\,$mK when the beam power is increased. Another reason could be the occurence of collisions involving doubly-excited Li*K* atom pairs, the rate of which increases with the excited state populations. The scattering potentials for these collisions are known to be of a long-range, as they scale with the internuclear separation as $1/R^5$~\cite{MarSad99}, whereas they scale as $1/R^6$ for collisions involving a singly-excited heteronuclear atom pair~\cite{DerJoh02}).

\section{Conclusions}
We have produced a dual-species magneto-optical trap for fermionic $^{6}$Li and $^{40}$K with large atom numbers. Two strategies have been applied in order to achieve this result. First, the dual-species MOT is placed in an ultra-high vacuum environment, being continuously loaded from cold atomic beams. The atomic beams originate from separate atom sources---a Zeeman slower for $^{6}$Li and a 2D-MOT for $^{40}$K---which both yield a large flux of cold atoms. Second, the homo- and heteronuclear collisions have been minimized by using small magnetic field gradients and low light powers in the repumping light. The atom loss in each MOT due to the presence of the other species decreases by only 4\% ($^{6}$Li) and 10\% ($^{40}$K) due to the heteronuclear collisions.

We have given a detailed description of the implemented apparatus, which we hope serves as a guideline for the construction of next generation experiments with fermionic $^{6}$Li and $^{40}$K.

The produced dual-species MOT represents the starting point for the production of a large-atom number quantum degenerate Fermi-Fermi mixture. The atoms trapped in the dual-species MOT have already been transferred into the magnetic trap and magnetically transported to the science chamber with large optical access and low background pressure. The large depth of magnetic traps as compared to optical traps allows for a large transfer efficiency, leading to smaller losses of atoms. In the science cell, the dual-species cloud will be evaporatively cooled in a plugged magnetic trap to quantum degeneracy and then transferred into an optical trap for investigation. 

\begin{acknowledgement}
The authors acknowledge support from ESF Euroquam (FerMix), SCALA, ANR FABIOLA, R\'egion Ile de France (IFRAF), ERC Ferlodim and Institut Universitaire de France. A.R. acknowledges funding from the German Federal Ministry of Education and Research and D.R.F. from Funda\c{c}\~{a}o para a Ci\^encia e Tecnologia (FCT) through grant SFRH/BD/68488/2010 and from Funda\c{c}\~{a}o Calouste Gulbenkian.

\end{acknowledgement}

\begin{appendix}

\section{Tapered Amplifier Mounts}

We developed compact support designs for our tapered amplifier chips, in order to minimize the costs of the laser sources of our experimental setup. The TAs are commercial semiconductor chips which are mounted on homemade compact mechanical supports with nearly no adjustable parts. The support designs allow for an easy installation process, which does not require any gluing or the help of micrometric translation stages for the alignment of the collimation optics, as that can be accomplished by free hand. Furthermore, the design minimizes the heat capacity of the support and the produced temperature gradients, allowing for a quick temperature stabilization that makes the TAs quickly operational after switch-on. The temperature stabilization is accomplished using a Peltier element (Roithner Lasertechnik GmbH, ref.~TEC1-12705T125) connected to a PID control circuit. The heat of the chip is dissipated via an aluminum base plate which is economically cooled by air rather than running water (the base plate reaches a maximum temperature of 28$^\circ$C for diode currents of 2A). 

\begin{figure}[h]
\centering
\includegraphics[width=8cm]{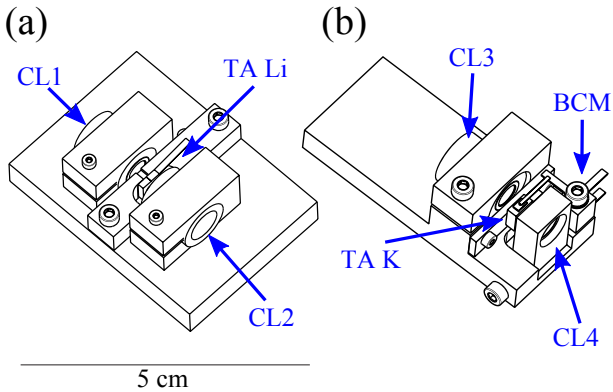}
\caption{(Color online) Sketch of the tapered amplifier supports for (a) Li and (b) K. In the figure, TA Li and TA K refer to the respective tapered amplifier chips, CL1, CL2, CL3 and CL4 to the (only longitudinally adjustable) collimation lens supports and BCM to the isolated mount for the blade connectors used to power the chip for K. The supports for the output collimation lenses are CL2 and CL4. }\label{TAmounts}
\end{figure}

The commercial TA chips are sold on small heat sinks which have different dimensions for the two different wavelengths. We thus had to design slightly different supports for the Li- and K-TAs, which are both schematically shown in Fig.~\ref{TAmounts}.

For lithium the semiconductor chip (Toptica, ref.~TA-670-0500-5) is delivered on a heat dissipation mount of type ``I''. It is placed between two axially aligned cylindrical lens tubes (CL1 and CL2 in Fig.~\ref{TAmounts} (a)), each of which containing an aspheric collimation lens of focal length 4.5\,mm (Thorlabs, ref.~C230TME-B). The support of the tubes and the chip are precisely machined such that the chip's output beam falls on the center of the respective collimation lens (CL2 in Fig.~\ref{TAmounts} (a)). The tubes are supported by cylindrically holed tightenable hinges in which they can move only longitudinally, along the direction of the amplified laser beam. This restriction of the tube's motion facilitates the alignment of the collimation lenses. The support design does not allow for a transverse alignment of the collimation lenses. Since this alignment is not very critical for the performance of the TA, we found it needless to allow this degree of freedom and relied on precise machining (possible imperfections could be compensated utilizing the mechanical play of the large attachment screw holes of the commercial heat sinks of the chips). When tightened by a screw, the hinges fix the position of the tubes. Since the tightening applies a force perpendicular to the longitudinal direction, it does not move the tubes along this (critical) direction. They might only move slightly along the transverse direction, which does not affect the final performance of the TA. 

For potassium, the semiconductor chip (Eagleyard, ref. EYP-TPA-0765-01500-3006-CMT03-0000) is delivered on a heat dissipation mount of type ``C''. Placing this mount between two hinges as for the case of lithium is less convenient since the heat dissipation mount has to be attached by a screw in the longitudinal direction which requires access from one side. Therefore one hinge is replaced by a rail which guides a parallelepipedically formed mount for the second (output) collimation lens (CL4 in Fig.~\ref{TAmounts} (b)). The motion of this mount is also fixed by tightening a screw applying forces perpendicular to the rail direction, which does not move the collimation lens along the critical longitudinal direction. For all our TAs, the positioning of the collimation lenses never had to be adjusted again once they were aligned. 

%
%
The commercial heat dissipation mount of the potassium chip is inconvenient for a simple powering of the chip. The very fragile gold wire, which has to be connected to the negative source of the current supply, has to be protected by a mechanical support before being connected to a cable. Therefore we soldered it to a blade connector that is fixed by an isolated plastic mount (BCM Fig.~\ref{TAmounts} (b)) and which is connected to the current supply. To avoid an overheating of the chip during the soldering process we permanently cooled the gold wire by blowing cold dry air from a spray can on it.

The output beams of the TA chips are astigmatic and thus require additional collimation. The choice of the collimation optics needs to be adapted to the specifications of the subsequent optical fiber, which in our case requests a collimated circular Gaussian beam of 2.2\,mm \linebreak 1/e$^2$-diameter for optimum coupling efficiency. The mode-matching was found optimum for a pair of lenses consisting of one spherical lens (with f=15\,cm for Li and f=4\,cm for K) and a cylindrical lens (with f=8\,cm for Li and f=2.54\,cm for K), which are placed outside the TA's housing. The cylindrical lenses are supported by rotatable mounts, in order to facilitate the mode-matching into the fibers. For all our TAs we achieve fiber-coupling efficiencies larger than 50\% (Li) and 60\% (K).

When injected with 20\,mW, the Li-TAs yield an output power of 500\,mW at 1\,A driving current and the K-TAs yield an output power of 1500\,mW at 2.5\,A driving current. In order to increase the lifetime of the chips, we limit the driving currents to smaller values and we switch the chips on only for periods of experimentation. When switched on, the TAs quickly reach a stable functioning (usually within 10\,min) due to the compactness of the mechanical support, which allows for a quick temperature stabilization.

\end{appendix}

%
\bibliographystyle{unsrt}
\bibliography{bibliography}
%

\end{document}